\documentclass[lettersize,journal]{IEEEtran}

\usepackage{graphicx}
\usepackage{epstopdf}
\usepackage[caption=false,font=footnotesize]{subfig}
\usepackage{cite}
\usepackage{color}
\usepackage{xcolor}
\usepackage[cmex10]{amsmath}
\usepackage{amssymb}
\usepackage{threeparttable}
\usepackage{mathtools}
\usepackage{hyperref}
\usepackage{algpseudocode}
\usepackage{multirow}
\usepackage[T1]{fontenc}
\usepackage[ruled]{algorithm2e}
\usepackage{booktabs}
\usepackage{stfloats}
\usepackage{tensor}
\usepackage{bbm}
\usepackage{multicol}
\usepackage{tikz}
\usepackage{cooltooltips}

\definecolor{mygreen}{rgb}{0.1,.6,0.1}
 
\newcommand{\minsum}{Min Sum}
\begin{document}

\title{Reconstruction-Computation-Quantization (RCQ): \\A Paradigm for Low Bit Width LDPC Decoding}

\author{
Linfang Wang~\IEEEmembership{Student Member,~IEEE,}  Caleb Terrill{Student Member,~IEEE,}  Maximilian Stark, Zongwang Li, Sean Chen, Chester Hulse, Calvin Kuo, Richard Wesel~\IEEEmembership{Fellow,~IEEE,}, Gerhard Bauch~\IEEEmembership{Fellow,~IEEE,},  Rekha Pitchumani

	\thanks{This paper was presented in part at IEEE GLOBECOM 2020,2021 \cite{Wang2020-RCQ,Terrill2021-ec}.}
	\thanks{This research is supported by National Science Foundation (NSF) grant CCF-1911166 Physical Optics Corporation (POC) and SA Photonics. Any opinions, findings, and conclusions or recommendations expressed in this material are those of the author(s) and do not necessarily reflect views of the NSF, POC, or SA.}
	\thanks{Linfang Wang,  Caleb Terrill, Sean Chen, Chester Hulse, Calvin Kuo and Richard Wesel  are with the Department of Electrical and Computer Engineering, University of California, Los Angeles, Los Angeles, CA, 90095 USA.  E-mail: \{lfwang, cterrill26, mistystory, chulse, calvinkuo, wesel\}@ucla.edu.}
\thanks{Zongwang Li and  Rekha Pitchumani are with Samsung Semiconductor Inc., San Jose, CA, 95134, United State. E-mail: \{zongwang.li, r.pitchumani\}@samsung.com.}
\thanks{ M. Stark and G. Bauch are with the Institute of Communications, Hamburg University of Technology, Hamburg, 21073, Germany. E-mail: \{maximilian.stark, bauch\}@tuhh.de.}}

\markboth{Journal of \LaTeX\ Class Files,~Vol.~14, No.~8, August~2021}%
{Shell \MakeLowercase{\textit{et al.}}: A Sample Article Using IEEEtran.cls for IEEE Journals}

\IEEEpubid{TCOM-TPS-21-1081.R1\$00.00~\copyright~2021 IEEE}

\maketitle

\begin{abstract}
This paper uses the reconstruction-computation-quantization (RCQ) paradigm to decode low-density parity-check (LDPC) codes.  RCQ facilitates dynamic non-uniform quantization to achieve good frame error rate (FER) performance with very low message precision.  For message-passing according to a flooding schedule, the RCQ parameters are designed by discrete density evolution.  Simulation results on an IEEE 802.11 LDPC code show that for 4-bit messages, a flooding \minsum RCQ decoder outperforms  table-lookup approaches such as information bottleneck (IB) or Min-IB decoding, with significantly fewer parameters to be stored.

Additionally, this paper introduces layer-specific RCQ, an extension of RCQ decoding for layered architectures.  Layer-specific RCQ uses layer-specific message representations to achieve the best possible FER performance.  For layer-specific RCQ, this paper proposes using layered discrete density evolution featuring hierarchical dynamic quantization (HDQ) to design parameters efficiently.

Finally, this paper studies field-programmable gate array (FPGA) implementations of RCQ decoders. Simulation results for a (9472, 8192) quasi-cyclic (QC) LDPC code show that a layered \minsum RCQ decoder with 3-bit messages achieves more than a $10\%$ reduction in LUTs and routed nets and more than a $6\%$ decrease in register usage while maintaining comparable decoding performance, compared to a 5-bit offset \minsum decoder.

\end{abstract}

\IEEEpeerreviewmaketitle

\section{Introduction}

\IEEEPARstart{L}{ow}-Density Parity-Check  (LDPC) codes \cite{GallagerPhD1963} have been implemented broadly, including in NAND flash systems and wireless communication systems. Message passing algorithms such as belief propagation (BP) and \minsum are utilized in LDPC decoders. In practice, decoders with low message bit widths are desired when considering the limited hardware resources such as area, routing capabilities, and power utilization of FPGAs or ASICs. Unfortunately, low bit width decoders with uniform quantizers typically suffer a large degradation in decoding performance\cite{-_Lee2005-MIMQBP}. On the other hand, the iterative decoders that allow for the dynamic growth of message magnitudes can achieve improved performance\cite{Zhang2014-ib}.

LDPC decoders that quantize messages  non-uniformly have gained attention because they provide excellent decoding performance with low bit width message representations. 
One family of non-uniform LDPC decoders use lookup tables (LUTs) to replace the mathematical operations in the check node (CN) unit  and/or the variable node (VN) unit. 
S. K. Planjery \emph{et al.} propose finite alphabet iterative decoders (FAIDs) for regular LDPC codes in \cite{Planjery2013-FAIDI,Declercq2013-FAIDII},  which optimize a \textit{single} LUT to describe VN input/output behavior. In \cite{Planjery2013-FAIDI} a FAID is  designed to tackle certain trapping sets and hence achieves a lower error floor than BP on the binary symmetric channel (BSC).  Xiao \emph{et al.} optimize the parameters of FAID using a recurrent quantized neural network (RQNN)\cite{Xiao2019-RNNFIAD,Xiao2020-RNNFAID}, and the simulation results show that RQNN-aided linear FAIDs are capable of surpassing floating-point BP in the waterfall region for regular LDPC codes.

Note that the size of the LUTs in \cite{Planjery2013-FAIDI,Declercq2013-FAIDII,Xiao2019-RNNFIAD,Xiao2020-RNNFAID} describing VN behavior are an exponential function with respect to node degree. Therefore, these FAIDs can only handle regular LDPC codes with small node degrees. For codes with large node degrees, Kurkoski {\em et al}. develop a mutual-information-maximization LUT (MIM-LUT) decoder in \cite{kurkoski2016-IB}, which decomposes a single LUT with multiple inputs into a series of concatenated $2\times1$ LUTs, each with two inputs and one output. This decomposition makes the number of LUTs linear with respect to node degree, thus significantly reducing the required memory.  The MIM-LUT decoder performs lookup operations at both the CNs and VNs. The 3-bit MIM-LUT decoder shows a better FER than floating-point BP over the additive white Gaussian noise (AWGN) channel. As the name suggests, the individual $2\times1$ LUTs are designed to maximize mutual information\cite{Kurkoski2014-QuanDMC}.   
\IEEEpubidadjcol
Lewandowsky \emph{et al.} use the information bottleneck (IB) machine learning method to design LUTs and propose an IB decoder for regular LDPC codes. As with MIM-LUT, IB decoders also use $2\times1$ LUTs at both CNs and VNs. Stark \emph{et al.} extend the IB decoding structure to support irregular LDPC codes through the technique of message alignment \cite{Stark2018-IBMA,Stark2021-ai}. 
The IB decoder shows an excellent performance on a 5G LDPC code\cite{Stark2020-IB5G,Stark2020-IBjournal}. In order to reduce the memory requirement for LUTs, Meidlinger \emph{et al.} propose the Min-IB decoder, which replaces the LUTs at CNs with label-based min operation \cite{Meidlinger2015-MIMIB,Meidlinger2017-MINIBIRR,Meidlinger2020-MINIBIRR,Ghanaatian2018-MINIB-588}.

Because the decoding requires only simple lookup operations, the LUT-based decoders deliver high throughput. However, the LUT-based decoders require significant memory resources when the LDPC code has large degree nodes and/or the decoder has a large predefined maximum decoding iteration time, where each iteration requires its own LUTs. The huge memory requirement for numerous large LUTs prevents these decoders from being viable options when hardware resources are constrained to a limited number of LUTs.

Lee \emph{et al.}\cite{-_Lee2005-MIMQBP} propose the mutual information maximization quantized belief propagation (MIM-QBP) decoder which circumvents the memory problem by designing non-uniform quantizers and reconstruction mappings at the nodes. Both VN and CN operations are simple mappings and fixed point additions in MIM-QBP. He \emph{et al.} in \cite{He2019-MIMQBP} show how to systematically design the MIM-QBP parameters for quantizers and reconstruction modules. Wang {\em et al.} further generalize the MIM-QBP structure and propose a reconstruction-computation-quantization (RCQ) paradigm \cite{Wang2020-RCQ} which allows CNs to implement either the min or boxplus operation.

All of the papers discussed above focus on decoders that use the flooding schedule. The flooding schedule can be preferable when the code length is short. However, in many practical settings such as coding for storage devices where  LDPC codes with long block lengths are selected, the flooding schedule requires an unrealistic amount of parallel  computation for some typical hardware implementations. Layered decoding\cite{shuffled}, on the other hand, balances parallel computations and resource utilization for a  hardware-friendly implementation that also reduces the number of iterations as compared to a flooding implementation for the same LDPC code.

\subsection{Contributions}
As a primary contribution, this work extends our previous work on RCQ \cite{Wang2020-RCQ} to provide dynamic quantization that changes with each layer of a layered LDPC decoder, as is commonly used with a protograph-based LDPC code.  The original RCQ approach \cite{Wang2020-RCQ}, which uses the same quantizers and reconstructions for all layers of an iteration, suffers from FER degradation and a high average number of iterations when applied to a layered decoding structure.  The novelty and contributions in this paper are summarized as follows:
\begin{itemize}
    \item \textit{Layer-specific RCQ Decoding structure.} This paper proposes the layer-specific RCQ decoding structure. The main difference between the original RCQ of  \cite{Wang2020-RCQ} and the layer-specific RCQ decoder is that layer-specific RCQ designs quantizers and reconstructions for each layer of each iteration. The layer-specific RCQ decoder provides better FER performance and requires a smaller number of iterations than the original RCQ structure with the same bit width. This improvement comes at the cost of an increase in the number of parameters that need to be stored in the hardware.  
    \item \textit{layer-specific RCQ Parameter Design.}
    This work uses layer-specific discrete density evolution featuring hierarchical dynamic quantization (HDQ) to design the layer-specific RCQ parameters.  We refer to this design approach as layer-specific HDQ discrete density evolution. For each layer of each iteration, layer-specific HDQ discrete density evolution separately computes the PMF of the messages. HDQ designs distinct quantizers and reconstructions for each layer of each iteration.
    \item \textit{FPGA-based RCQ Implementations.} This paper presents the Lookup Method, the Broadcast Method and the Dribble Method, as alternatives to distribute RCQ parameters efficiently in an FPGA. This paper verifies the practical resource needs of RCQ through an FPGA implementation of an RCQ decoder using the Broadcast method.  Simulation results for a (9472, 8192) quasi-cyclic (QC) LDPC code show that a layer-specific \minsum RCQ decoder with 3-bit messages achieves a more than $10\%$ reduction in LUTs and routed nets and more than a $6\%$ reduction in register usage while maintaining comparable decoding performance, compared to a standard offset \minsum decoder with 5-bit messages.
 \end{itemize}
\subsection{Organization}
The remainder of this paper is organized as follows: Sec. \ref{sec: RCQ-decoding-stru} introduces the RCQ decoding structure and presents an FPGA implementation of an RCQ decoder. Sec. \ref{Sec: HDQ} describes HDQ, which is used for channel observation quantization and RCQ parameter design. Sec. \ref{sec: layered} shows the design of the layer-specific RCQ decoder. Sec.  \ref{sec: simulation result} presents simulation results including FER and hardware resource requirements. Sec. \ref{sec: conclusion} concludes our work.

\section{The RCQ Decoding Structure}\label{sec: RCQ-decoding-stru}

The updating procedure of message passing algorithms contains two steps: 1) computation of the output message, 2) communication of the message to the neighboring node. To reduce the complexity of message passing, the computed message is often quantized  before being passed  to the neighboring node.  We refer to the computed messages as the \textit{internal messages}, and communicated messages passed over the edges of the Tanner graph as  \textit{external messages}.

When external messages are produced by a uniform quantizer, low bit width external messages can result in an early error floor \cite{Zhang_q1quasi}.
Thorpe \emph{et al.} introduced a non-uniform quantizer in \cite{-_Lee2005-MIMQBP}.  Their decoder adds a non-uniform quantizer and a reconstruction mapping to the output and input of the hardware implementation of each node unit. 
This approach delivers excellent decoding performance even with a low external bit width.  The RCQ decoder \cite{Wang2020-RCQ} can be seen as a generalization of the decoder introduced in \cite{-_Lee2005-MIMQBP}.

In this section, we provide detailed descriptions of the RCQ decoding structure. Three FPGA implementation methods for realizing the RCQ functionality are also presented.
\begin{figure}
    \centering
    \includegraphics[width=20pc]{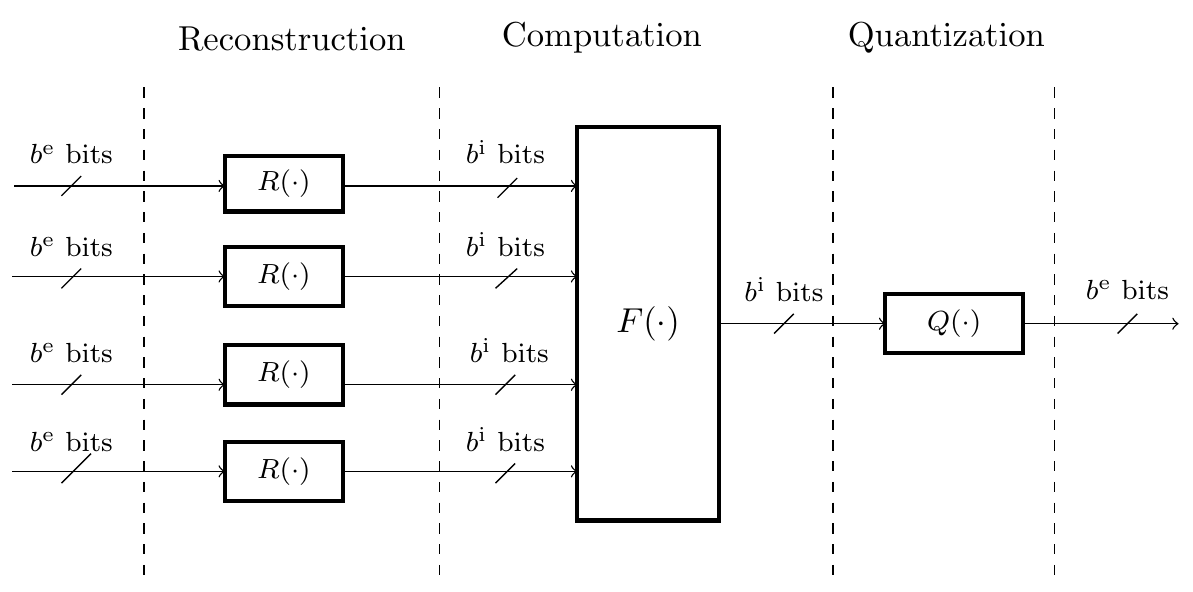}
    \caption{Illustration of a generalized RCQ unit which consists of three modules: \emph{Reconstruction} that maps a $b^{\text{e}}$-bit value to a $b^{\text{i}}$-bit value, \emph{Computation} that performs arithmetic operations, and \emph{Quantization} that quantizes a $b^{\text{i}}$-bit value to a $b^{\text{e}}$-bit value.}
    \label{fig: RCQ}
\end{figure}
\subsection{Generalized RCQ Unit}\label{sec: g-rcq}

 A generalized RCQ unit as  shown in Fig. \ref{fig: RCQ} consists of the following three modules:

\subsubsection{Reconstruction Module}\label{sec: recons}The reconstruction module applies a reconstruction function $R(\cdot)$ to each incoming $b^{\text{e}}$-bit external message to produce a $b^{\text{i}}$-bit internal message, where $b^\text{i}>b^\text{e}$. We denote the bit width of CN and VN internal message by $b^{\text{i,c}}$ and $b^{\text{i,v}}$, respectively. For the flooding-scheduled RCQ decoder, $R(\cdot)$ is iteration-specific and we use $R^{(t)}_\text{c}(\cdot)$ and $R^{(t)}_\text{v}(\cdot)$ to represent the reconstruction of check and variable node messages at iteration $t$, respectively. In the layer-specific RCQ decoder, $R(\cdot)$ uses distinct parameters for each layer in each iteration.  We use $R_\text{c}^{(t,r)}(\cdot)$ and $R_\text{v}^{(t,r)}(\cdot)$ to represent the the reconstruction of check and variable node messages at  layer $r$ of iteration $t$, respectively. The reconstruction functions are mappings of the input external messages to log-likelihood ratios (LLR) that will be used by the node.  In this paper, these mappings are systematically designed by HDQ discrete density evolution, which will be introduced in a later section.

For a quantizer $Q(\cdot)$ that is symmetric, an external message $d\in\mathbb{F}_2^{b^\text{e}}$ can be represented as $[d^{\text{MSB}}\ \tilde{d}]$, where $d^{\text{MSB}}\in\{0,1\}$ indicates sign and $\Tilde{d}\in\mathbb{F}_2^{b^\text{e}-1}$ corresponds to magnitude. We define the magnitude reconstruction function $R^*(\cdot):\mathbb{F}_2^{b^\text{e}-1}\rightarrow\mathbb{F}_2^{b^\text{i}-1}$, which maps the magnitude of external message, $\tilde{d}$, to the magnitude of internal message. Without loss of generality,  we restrict our attention to monotonic reconstruction functions so that
\begin{align}
    R^*(\tilde{d}_1)>R^*(\tilde{d}_2)>0,\quad \text{for }\tilde{d}_1>\tilde{d}_2,\label{r_mono}
\end{align}
where $\tilde{d}_1$, $\tilde{d}_2\in \mathbb{F}_2^{b^\text{e}-1}$. The reconstruction $R(d)$ can be expressed by
$R(d)=\left[d^{\text{MSB}}\ \ R^*(\tilde{d})\right]$.
Under the assumption of a symmetric channel, we have $R([0\ \Tilde{d}])=-R([1\ \Tilde{d}])$.
\subsubsection{Computation Module}
The computation module \texorpdfstring{$F(\cdot)$}{Lg} uses the $b^\text{\text{i}}$-bit outputs of the reconstruction module to compute a $b^\text{i}$-bit internal message for the CN or VN output. We denote the computation module implemented in CNs and VNs by $F_\text{c}$ and $F_\text{v}$, respectively. An RCQ decoder implementing the min operation at the CN yields a \minsum (ms) RCQ decoder. If an RCQ decoder implements belief propagation (bp) via the \textit{boxplus} operation, the decoder is called \textit{bpRCQ}. The computation module, $F_\mathrm{v}$, in the VNs is addition for both bpRCQ and msRCQ decoders.


\subsubsection{Quantization Module} The quantization module \texorpdfstring{$Q(\cdot)$}{Lg} quantizes the $b^{\text{i}}$-bit internal message to produce a $b^{\text{e}}$-bit external message.
Under the assumption of a symmetric channel, we use a symmetric quantizer that features sign information and a magnitude quantizer $Q^*(\cdot)$.
The magnitude quantizer selects one of $2^{b^\text{e}-1}-1$ possible indexes using the threshold values $\{\tau_0,\tau_1,...,\tau_{\text{max}}\}$, where  $\tau_j\in\mathbb{F}_2^{b^{\mathrm{i}}}$ for $j\in\{0,1,...,2^{b^\text{e}-1}-2\}$ and $\tau_{\text{max}}$ is $\tau_{j_{\text{max}}}$ for $j_{\text{max}} = 2^{b^\text{e}-1}-2$.  We also require
\begin{align}
    {\tau}_i>{\tau}_j>0,\quad i>j.
\end{align}
Given an internal message $h\in\mathbb{F}_2^{b^{\mathrm{i}}}$, which can be decomposed into sign part $h^{\text{MSB}}$ and magnitude part $\tilde{h}$, $Q^*(\tilde{h})\in \mathbb{F}_2^{b^{\mathrm{e}}-1}$ is defined by:
\begin{align}
   {{Q^*(\tilde{h})}}=\left\{\begin{matrix}
 0,& \tilde{{h}} \leq {\tau}_0 \\ 
 j, &  {\tau}_{j-1} <  \tilde{{h}} \leq {\tau}_{j} \\
 2^{b^{\text{e}}-1}-1, & \tilde{h} >  {\tau}_{\text{max}}
\end{matrix}\quad,\right.\label{equ: quantization_mag}
\end{align}
where $0<j\le j_{\text{max}}$. Therefore, $Q(h)$ is defined by
$Q(h)=[h^{\text{MSB}}\ Q^*(\tilde{h})]$.
The super/subscripts introduced for $R(\cdot)$ also apply to $Q(\cdot)$. 
\subsection{Bit Width of RCQ decoder}
 The three tuple $(b^\text{e},b^{\text{i,c}}, b^{\text{i,v}})$ represents the precision of messages in a RCQ decoder. 
For the \emph{msRCQ} decoder, it is sufficient to use only the pair $(b^\text{e},b^{\text{i,v}})$ because $b^{\text{i,c}}=b^\text{e}$, we simply denote $b^{\text{i,v}}$ by $b^{\text{v}}$.   The CN min operation computes the XOR of the sign bits and finds the minimum of the extrinsic magnitudes. For a symmetric channel, the min operation can be computed by manipulating the external messages, because the external message delivers the  \emph{relative LLR meaning}  of reconstructed values.  Since we only use external messages to perform the min operation, $R^\text{c}(\cdot)$ and $Q^\text{c}(\cdot)$ are not needed for the {msRCQ decoder}. Finally, we use $\infty$ to denote a floating point representation. 
\subsection{FPGA Implementation for RCQ}
The RCQ FPGA decoder may be viewed as a modification to existing hardware decoders based on the BP or MS decoder algorithms, which have been studied extensively\cite{Zhang_undated-es,Sadek2016-lz,Liu2017-ho,Anantharaman2019-db}. 
The RCQ decoders require extra  $Q(\cdot)$ and $R(\cdot)$ functions to quantize and reconstruct message magnitudes. 
To implement  $Q(\cdot)$ and $R(\cdot)$ functions, we have devised the \emph{Lookup}, \emph{Broadcast}, and \emph{Dribble} methods. 
These three approaches are functionally identical, but differ in the way that the parameters needed for the $Q(\cdot)$ and $R(\cdot)$ operations are communicated to the nodes. 

\subsubsection{Lookup Method}
The quantization and reconstruction functions simply map an input message to an output message.  Thus, a simple implementation uses lookup tables implemented using read-only memories (ROMs) to implement all these mappings.
The $Q(\cdot)$ and $R(\cdot)$ functions in every VN require their own ROMs, implemented using block RAMs. 
Because $Q(\cdot)$ and $R(\cdot)$ change with respect to different iterations and/or layers, one potential drawback of the Lookup method is a large block RAM requirement.

\begin{figure}[t]
    \centering
      \subfloat[\label{recons_FPGA}]{%
        \includegraphics[width=.5\linewidth]{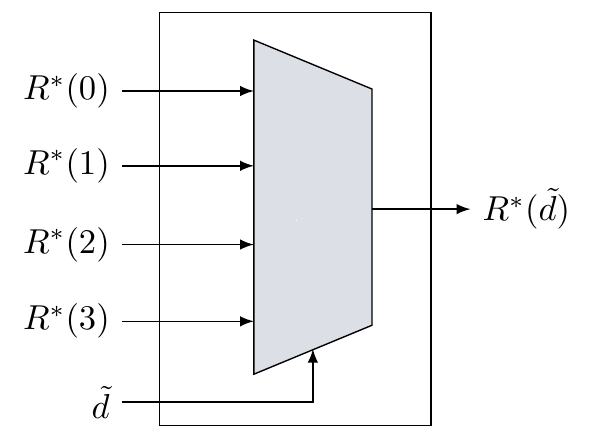}}
  \subfloat[\label{quan_FPGA}]{%
       \includegraphics[width=.5\linewidth]{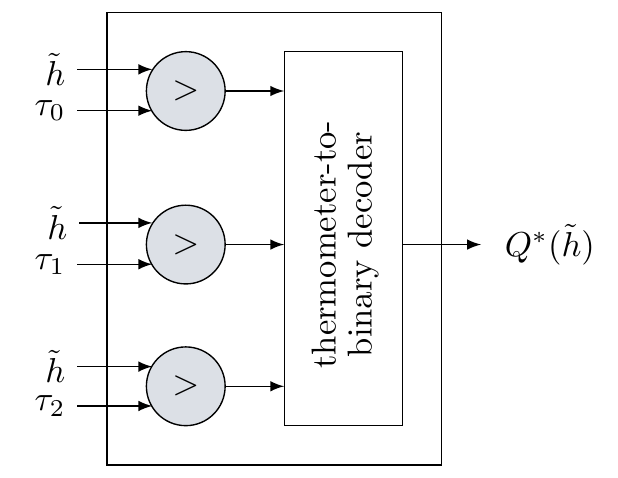}}
  \\
  \subfloat[\label{t2b_FPGA}]{%
       \includegraphics[width=.4\linewidth]{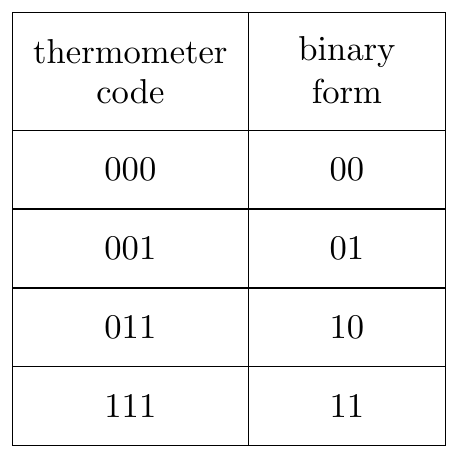}}
  \caption{msRCQ magnitude reconstruction module (a) and magnitude quantization module (b). In FPGA, magnitude reconstruction module is realized by a multiplexer, and  magnitude quantization is realized by comparison functions and  a thermometer-to-binary decoder which realizes the mapping relationship shown in (c).}
  \label{fig: FPGA}
\end{figure}

\subsubsection{Broadcast Method}
The Broadcast method provides a scheme where all RCQ parameters are stored centrally in a control unit, instead of being stored in each VN.  
Each VN only takes in the $Q(\cdot)$ and $R(\cdot)$ parameters necessary for decoding the current iteration and layer, and use logic to perform their respective operations. Fig. \ref{fig: FPGA} shows an implementation for a 3-bit RCQ, which uses mere 2 bits for magnitude reconstruction and quantization.
The $2$-bit magnitude reconstruction module  is realized by a $4\times1$ multiplexer. The $2$-bit magnitude quantization consists of two steps, first a  thermometer code\cite{Ajanya2018-uy}, where the contiguous ones are analogous to mercury in a thermometer, is generated by comparing the input with all thresholds, and then the thermometer code is converted to the  $2$-bit binary form by using a thermometer-to-binary decoder, which realizes the mapping relationship in Fig. \ref{t2b_FPGA}. 
Two block RAMS are required in the control unit for the thresholds and reconstruction values. Small LUTs in each VN implement the $Q(\cdot)$ and $R(\cdot)$ functions. The main penalty of the Broadcast method is the additional wiring necessary to route the RCQ parameters from the central control unit to the VNs.


\subsubsection{Dribble Method}
The Dribble method attempts to reduce the number of long wires required by the Broadcast method. Registers in the VNs save the current thresholds and reconstruction values necessary for the $Q(\cdot)$ and $R(\cdot)$ functions. Once again, quantization and reconstruction can be implemented using the logic in Fig. \ref{fig: FPGA}. When a new set of parameters is required, the bits are transferred (dribbled) one by one or in small batches from the control unit to the VN unit registers. Just as in the Broadcast method, two extra block RAMs and logic for the $Q(\cdot)$ and $R(\cdot)$ functions are required. 
The penalty of the Dribble method comes with the extra usage of registers in the VN units.

We have implemented all methods and explored their  resource utilization in \cite{Terrill2021-ec}.

\section{Hierarchical Dynamic Quantization (HDQ) }\label{Sec: HDQ}
This section introduces the HDQ algorithm, a non-uniform quantization scheme that this paper uses both for quantization of channel observations and for quantization of internal messages by RCQ. Our results show, for example, that HDQ quantization of AWGN channel observations achieves performance similar to the optimal dynamic programming quantizer of \cite{Kurkoski2014-QuanDMC} for the binary input AWGN channel, with much lower computational complexity. 

\subsection{Motivation}
The quantizer plays an important role in RCQ decoder design. First, the channel observation is quantized as the input to the decoder. This section explores how to use HDQ to quantize the channel observations.  Second, the parameters of $R(\cdot)$ and $Q(\cdot)$ are also designed by quantizing external messages according to their probability mass function (PMF) as determined by discrete density evolution.  The use of HDQ to quantize internal messages is described in Section \ref{sec: layered}. 

The HDQ approach designs a quantizer that maximizes mutual information in a greedy or progressive fashion.  Quantizers aiming to maximize mutual information are widely used in non-uniform quantization design\cite{He2019-MIMQBP,Wang2020-RCQ,Lewandowsky2018-IBRegular,Stark2018-IBMA,Stark2020-IB5G,Meidlinger2015-MIMIB,Meidlinger2020-MINIBIRR,Meidlinger2017-MINIBIRR,Stark2020-IBjournal,Ghanaatian2018-MINIB-588, nathan-hdq, jiadong_softinfo_conf, jiadong_softinfo_jour}.
Due to the interest of this paper, the cardinality of quantizer output is restricted to $2^b$, i.e., this paper seeks $b$-bit quantizers.
Kurkoski and Yagi \cite{Tal2011-QuanVardy} proposed a dynamic programming method to find an {optimal} quantizer that maximizes mutual information for a binary input discrete memoryless channel (BI-DMC) whose outputs are from an alphabet with cardinality $B$,  with complexity $\mathcal{O}(B^3)$.
The dynamic programming method of \cite{Kurkoski2014-QuanDMC} finds the optimal quantization, but the approach becomes impractical when $B$ is large.

In order to quantize the outputs for a channel with large cardinality $B$ when constructing polar codes,  Tal and Vardy devised a sub-optimal greedy quantization algorithm with complexity  $\mathcal{O}(B\log(B))$ \cite{Tal2011-QuanVardy}. In \cite{Lewandowsky2018-IBRegular}, Lewandowsky \textit{et al.} proposed the modified Sequential Information Bottleneck (mSIB) algorithm to design the channel quantizer and LUTs for LDPC decoders. mSIB is also a sub-optimal quantization technique with complexity $\mathcal{O}(aB)$, where $a$ is the number of trials. As a machine learning algorithm, multiple trials are required for good results with mSIB.  Typical values of $a$ range, for example, from 15 to 70.

HDQ is proposed in \cite{Wang2020-RCQ} as an efficient $b$-bit quantization algorithm for the symmetric BI-DMC with complexity $\mathcal{O}\left(\frac{2^b}{\log(\gamma)}\log(B)\right)$. HDQ has less complexity than mSIB and also the Tal-Vardy algorithm. This section reviews the HDQ using symmetric binary input AWGN channel as an example. As an improvement to the HDQ of \cite{Wang2020-RCQ}, sequential threshold search is replaced with golden section search\cite{Kiefer1953-pc}.

\subsection{The HDQ Algorithm}
Let the encoded bit $x\in\{0,1\}$ be modulated by Binary Phase Shift Keying (BPSK) and transmitted over an AWGN channel. The modulated BPSK signal is represented as $s(x)=-2x+1$. We denote the channel observation at the receiver by $y$ where
\begin{align}
    y = s(x) +z,
\end{align}
and $z\sim \mathcal{N}(0,\sigma^2)$. The joint probability density function of $x$ and $y$, $f(x,y;\sigma)$, is:
\begin{align}
    f(x,y;\sigma) = \frac{1}{2\sqrt{2\pi \sigma^2}}e^{-\frac{(y-s(x))^2}{2\sigma^2}}.
\end{align}

HDQ seeks an $b$-bit quantization of the continuous channel output $y$, as in \cite{jiadong_softinfo_conf}.  In practice, often $y$ is first quantized into $B$ values using high-precision uniform quantization where $B\gg 2^b$, i.e., analog-to-digital (A/D) conversion.   Let $W$ be the result of the A/D output, where $W\in \cal W $ and  ${\cal W} =\{0,1,...,B-1\}$. 
The alphabet of $B$ channel outputs from the A/D converter is then subjected to further non-uniform quantization resulting in a quantization alphabet of $2^b$ values. We use $D$ to represent the non-uniform quantizer output, which is comprised of the $b$ bits $ D=[D_1,...,D_b]$.  HDQ aims to maximize the mutual information between $X$ and $D$.

For the symmetric binary input AWGN channel, a larger index $w$ implies a larger LLR, i.e.:
\begin{align}\label{equ: inq}
    \log \frac{P_{W|X}(i|0)}{P_{W|X}(i|1)}<\log \frac{P_{W|X}(j|0)}{P_{W|X}(j|1)},~ \forall i<j.
\end{align}
Based on Lemma 3 in \cite{Kurkoski2014-QuanDMC}, any binary-input discrete memoryless channel that satisfies \eqref{equ: inq} has an optimal $b$-bit quantizer that is determined by $2^b-1$ boundaries, which can be identified by their corresponding index values.  Denote the $2^{b}-1$ index thresholds by $\{\xi_{1}$, $\xi_{2},..., \xi_{2^{b}-1} \} \subset {\cal W}$. Unlike the dynamic programming algorithm\cite{Kurkoski2014-QuanDMC}, which optimizes boundaries jointly, HDQ \textit{sequentially} finds thresholds according to \textit{bit level}, similar to the progressive quantization in \cite{nathan-hdq}.

The general $b$-bit HDQ approach is as follows:

\begin{enumerate}
    \item We assume an initial high-precision uniform quantizer.  For this case, set the extreme index thresholds  $\xi_0=0$ and $\xi_{2^b}=B-1$, which are the minimum and maximum outputs of the uniform quantization.  
    \item The index threshold $\xi_{2^{(b-1)}}$ is selected as follows to determine the bit level 0:
    \begin {align}
    \xi_{2^{(b-1)}} = \arg \max_{\xi_0<\xi<\xi_{2^b}} I(X;D_1)\, , 
    \end{align}
where 
\begin{equation} 
D_1 = \mathbbm{1}(W\geq \xi_2^{(b-1)}).
\end{equation}

\item The index thresholds  $\xi_{2^{(b-2)}}$ and $\xi_{3*2^{(b-2)}}$ are selected as follows to determine bit level 1:
    \begin {align}
    \xi_{2^{(b-2)}} &= \arg \max_{\xi_0<\xi<\xi_{2^{b-1}}} I(X;D_2| D_1=0), \\ 
    \xi_{3*2^{(b-2)}} &= \arg \max_{\xi_{2^{b-1}}<\xi<\xi_{2^b}} I(X;D_2| D_1=1)\, ,
    \end{align}
    and 
\begin{equation} 
D_2 = \begin{cases} 
\mathbbm{1}(W \geq \xi_{2^{(b-2)}}) & \text{if } D_1=0 \\
\mathbbm{1}(W \geq \xi_{3*2^{(b-2)}}) & \text{if } D_1=1 \\
\end{cases}.
\end{equation}

\item In the general case, when the thresholds for $k$ previous quantization bits have been determined, $2^k$ thresholds $\{\xi_{(j+0.5)2^{b-k}}, j=0,..,2^{k}-1\}$ must be selected to determine the next quantization bit.  Each threshold maximizes $I(X;D_{k+1}|D_k=d_k, \ldots,D_1=d_1 )$ for a specific result for the $k$ previous quantization bits. 
\end{enumerate}



HDQ provides the $2^b -1$ index thresholds $\{\xi_1, \ldots, \xi_{2^b-1} \}$. For channel quantization, the index thresholds can be mapped to channel outputs. For the RCQ decoding, the messages are LLR values, the LLR magnitude thresholds $\{\tau_0,...,\tau_{2^{b-1}-2}\}$ are calculated from the index thresholds $\{\xi_{2^{b-1}+1}, \ldots ,\xi_{2^b-1} \}$ as follows:
\begin{align}
\tau_i =~ \log\frac{P_{W|X}(\xi_{1+i+2^{b-1}}|0)}{P_{W|X}(\xi_{1+i+2^{b-1}}|1)}, i=0,1,..,2^{b-1}-2.
\end{align}

HDQ also provides the joint probability between code bit $X$ and quantized message $D$, $P(X,D)$. The magnitude reconstruction function  $R^*(\cdot)$ is computed as follows: 
\begin{align}
    R^*(d) = \log\frac{P_{XT}(0,d+2^{b-1})}{P_{XT}(1,d+2^{b-1})},~d=0,1,...,2^{b-1}-1.
    \label{equ: recon_hdq}
\end{align}

\begin{figure}
	\centering
	\includegraphics[width=13pc]{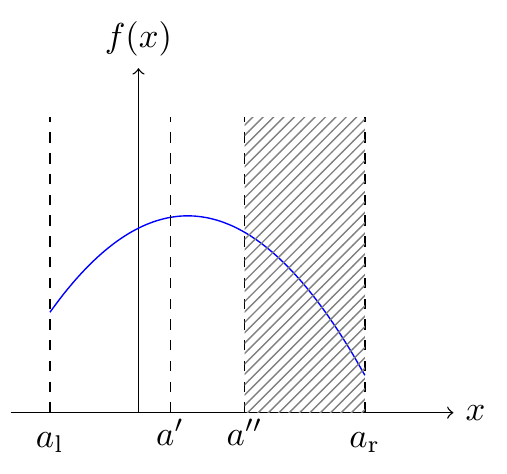}
	\caption{ Illustration of one iteration of golden-section search for finding maximum point of $f(x)$ in the interval $[a_\text{l},a_\text{r}]$. $a'=a_{\text{r}}-\frac{a_\text{r}-a_\text{l}}{\gamma}$ and $a''=a_{\text{l}}+\frac{a_\text{r}-a_\text{l}}{\gamma}$. Because $f(a'')<f(a')$, $[a'',a_\text{r}]$ is truncated and $[a_\text{l},a'']$ becomes the new search interval for the next iteration. 
	}
	\label{fig: gss}
\end{figure}

\subsection{Golden-Section Search and Complexity Analysis}\label{Sec: GSS}
After $k$ stages of HDQ, there are $2^k$ quantization regions each specified by their leftmost and rightmost indices $\xi_{\ell}$ and $\xi_r$.  The next stage finds a new threshold $\xi^*$ for each of these $2^k$ regions.   Each $\xi^*$ is selected to maximize a conditional mutual information as follows:
    \begin{equation}
    \xi^* = \arg \max_{\xi_{\ell}<\xi<\xi_r}  I(\xi),
    \label{equ: optimzation}
    \end{equation}
where 
\begin{align}
      I(\xi) &= I\left(X;D_{k+1}(\xi)|D_1=d_1, \ldots,D_k=d_k \right)\label{equ: I_xi} \\
    &= \sum_{x,d_{k+1}} P\left(x,d_{k+1}(\xi)|d_1^k\right) \log  \frac{P(d_{k+1}(\xi)|x,d_1^k) }{P(d_{k+1}(\xi)|d_1^k) }
\end{align}
for the binary $k$-tuple $d_1^k=d_1, \ldots, d_k$  that defines $(\xi_{\ell}, \xi_r)$. The probability $P\left(x,d_{k+1}(\xi)|d_1^k\right)$ is defined as follows:

\begin{equation}
    P\left(x,d_{k+1}(\xi)|d_1^k\right) = 
    \begin{cases}
    \frac{\sum_{w=\xi_l}^{\xi} P_{XW}(x,w)}{\sum_{w=\xi_l}^{\xi_r} P_{W}(w)} & d_{k+1}=0\\
    \frac{\sum_{w=\xi+1}^{\xi_r} P_{XW}(x,w)}{\sum_{w=\xi_l}^{\xi_r} P_{W}(w)} & d_{k+1}=1\\
    \end{cases}.
\end{equation}

Because $I(\xi)$ is concave in $\xi$, the local maximum can be found using the golden section search \cite{Kiefer1953-pc}, a simple but robust technique to find extreme point of a unimodal function by successively narrowing the range of values on a specified interval. Specifically, Fig. \ref{fig: gss} illustrates one iteration of golden-section search for finding maximum point of $f(x)$ in the interval $[a_\text{l},a_{\text{r}}]$. First, find $a'=a_{\text{r}}-\frac{a_\text{r}-a_\text{l}}{\gamma}$ and $a''=a_{\text{l}}+\frac{a_\text{r}-a_\text{l}}{\gamma}$, where $\gamma = \frac{\sqrt{5}+1}{2}$. Because $f(a'')<f(a')$, which suggests that the maximum point lies in $[a_\text{l},a'']$, the interval $[a'',a_\text{r}]$ is truncated and  $[a_\text{l},a'']$ is updated as the next round search interval. Further details of golden-section search can be found in \cite{Kiefer1953-pc}. When using the golden-section search to find all $2^b -1$ thresholds for the $b$-bit HDQ, $I(\xi)$ will be computed using \eqref{equ: I_xi} a number of times that is proportional to: 
\begin{align}
    &\log_{\gamma}(B)+\sum_{i=1}^{2^1}\log_{\gamma}(B_{2,i})+...+\sum_{i=1}^{2^{b-1}}\log_{\gamma}(B_{b,i}),\\
    \leq&\log_{\gamma}(B)+2\log_{\gamma}\left(\frac{B}{2}\right)+...+2^{b-1}\log_{\gamma}\left(\frac{B}{2^{b-1}}\right)\\=& \frac{2^b}{\log(\gamma)}\log(B).
\end{align}
 $B_{j,i}$ is the $i^{th}$ interval length in ${j-1}$ bit level quantization and $\sum_{i=1}^{2^{j-1}}B_{j,i}=B$. Therefore, a $b$-bit quantization on a  $B$-output channel using HDQ can be designed in $\mathcal{O}\left(\frac{2^b}{\log(\gamma)}\log(B)\right)$ time.

\begin{figure}
	\centering
	\includegraphics[width=16pc]{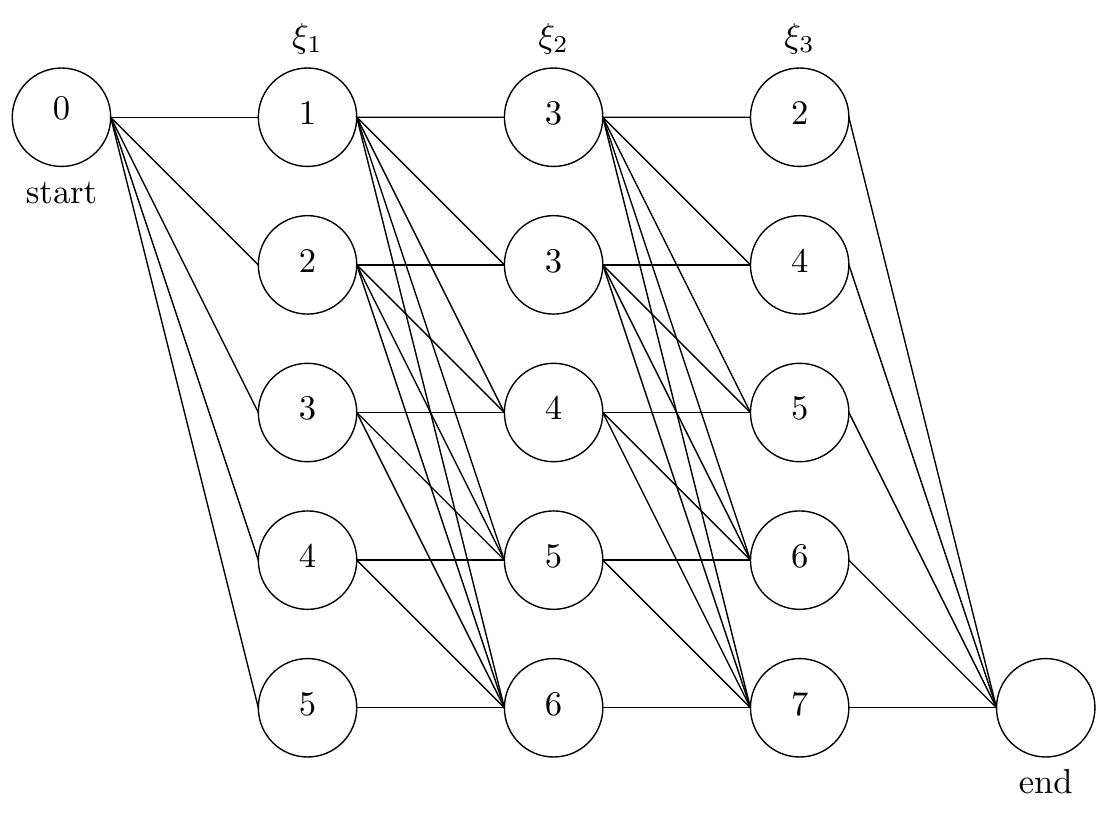}
	\caption{A trellis whose paths represent all 2-bit quantizers for a BI-DMC with 8 outputs. The vertices in column $i$ are possible values for $i^{th}$ threshold $\xi_i$. Each branch in the trellis identifies a quantization region.
	}
	\label{fig: trellis}
\end{figure}
\subsection{Comparing HDQ with Optimal Dynamic Programming}
This subsection provides an example contrasting HDQ with the dynamic programming solution. 
Following \cite{Kurkoski2014-QuanDMC}, Fig. \ref{fig: trellis} gives a trellis whose paths represent all 2-bit quantizers for a binary input DMC with 8 outputs. The outputs are indexed from 0 to 7 and satisfy \eqref{equ: inq}. The vertices in column $i$ are possible values for $\xi_i$, and each path represents a valid quantizer whose thresholds are determined by the vertices in each column. Each branch in the trellis identifies a quantization region. For example, the branch connecting vertex $\xi_0=0$ to vertex $\xi_1=2$ specifies the leftmost quantization region as \{0,1\}, i.e., $\xi_{\ell} = 0$ and $\xi_r =1$.
\begin{figure}[t] 
    \centering
  \subfloat[\label{fig: quan_result}]{%
       \includegraphics[width=1\linewidth]{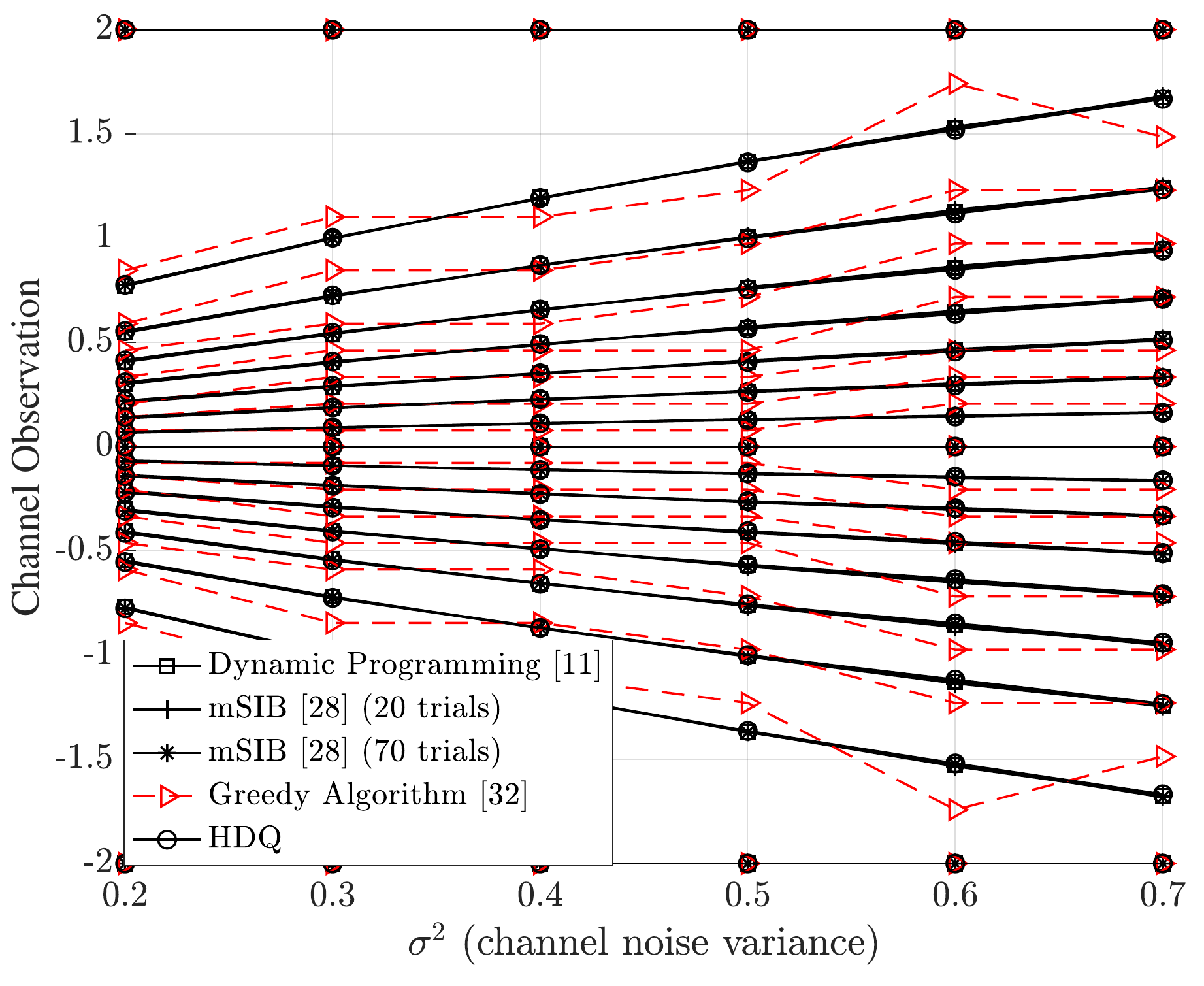}}
    \hfill
  \subfloat[\label{fig: mi_loss}]{%
        \includegraphics[width=1\linewidth]{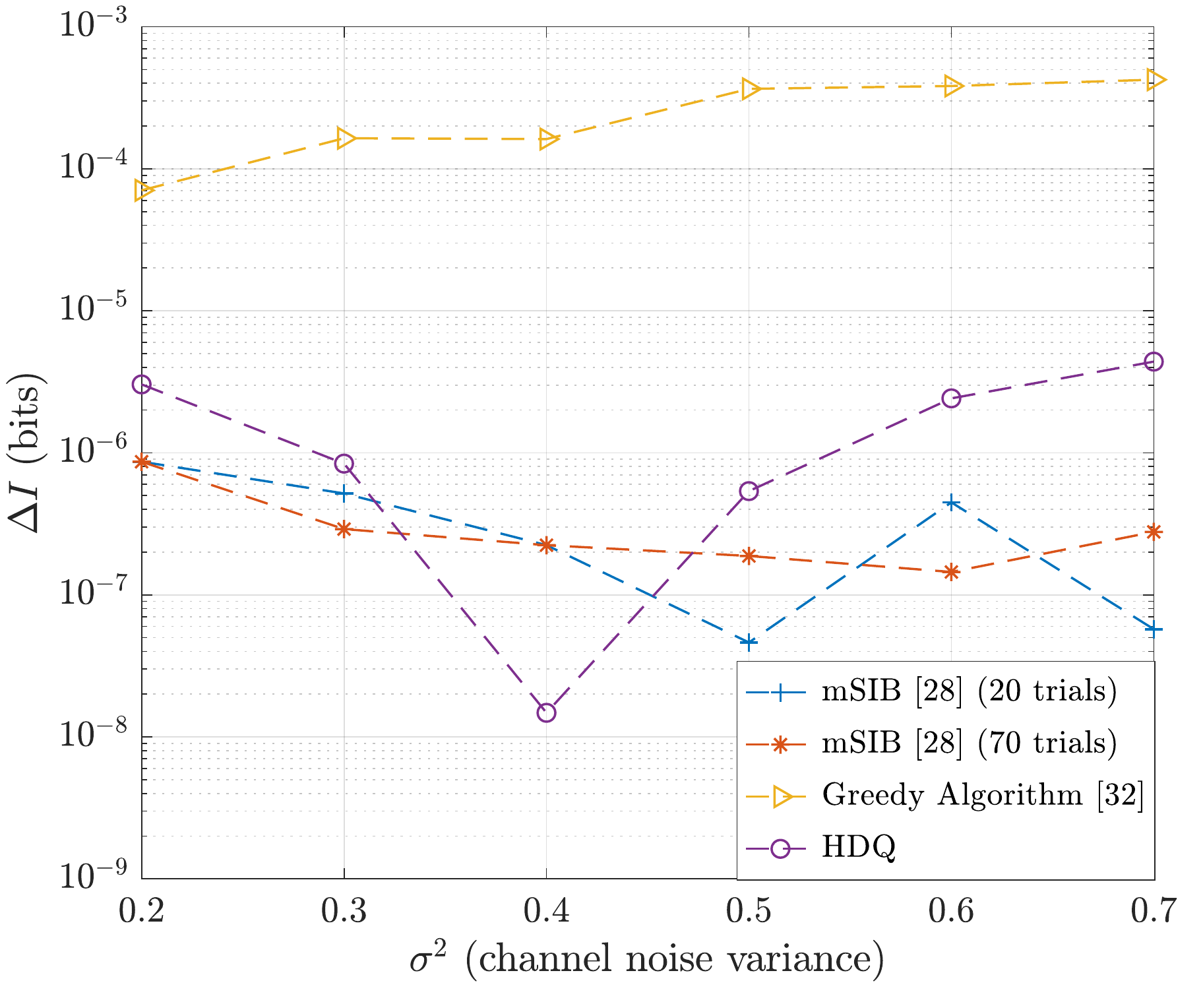}}
  \caption{Fig. (a): Quantization thresholds for dynamic programming, msIB, and HDQ on the BI-AWGNC as a function of $\sigma^2$ for $B=2000$. Fig. (b): Mutual information loss between each sub-optimal quantizer and optimal quantizer for BI-AWGNC as a function of $\sigma^2$ for $B=2000$. }
\end{figure}

 The dynamic programming algorithm determines vertices of all columns jointly, whereas HDQ identifies the vertices in a greedy way, by first finding the vertex in column 2 to maximize $I(X;D_1)$ and then vertices in column 1 and 4 to maximize $I(X;D_2|D_1=d_1)$.  Hence, the greedy approach of HDQ only searches part of trellis and therefore is sub-optimal. However, our simulations show that HDQ finds the quantizer that perform closely to  the optimal one.

\subsection{Simulation Result}
This section provides simulation results for quantizing symmetric binary input AWGN channel observations. The simulations compare HDQ to the optimal dynamic programming result as well as to two sub-optimal approaches: mSIB with 20 and 70 trials and the greedy quantization algorithm describe in \cite{Lewandowsky2018-IBRegular}. For all the quantization approaches, the channel observations are first quantized uniformly into $B=2000$ points between $-2$ and $2$. 

Fig. \ref{fig: quan_result} gives the thresholds as a function of $\sigma^2$ for HDQ, dynamic programming, mSIB with 20 and 70 trials, and  greedy quantization.  The quantization thresholds for HDQ, dynamic programming, and mSIB are indistinguishable in Fig.  \ref{fig: quan_result}.  HDQ has significantly lower complexity than both dynamic programming and mSIB. The thresholds for greedy quantization algorithm of \cite{Tal2011-QuanVardy} deviate noticeably from the thresholds found by the other approaches.

In order to quantify the performance of sub-optimal quantizers, we define $\Delta I$ as follows:
\begin{align}
    \Delta I = I^{\text{dp}}(X;D)-I^{\text{sub}}(X;D),
\end{align}
where $I^{\text{dp}}(X;D)$ and $I^{\text{sub}}(X;D)$ are the mutual information between code bit $X$ and quantized value $D$ as obtained by dynamic programming and sub-optimal quantizers, respectively. Fig. \ref{fig: mi_loss} plots $\Delta I$ as a function of $\sigma^2$ for each sub-optimal quantizer.  All three sub-optimal quantizers perform quite well with $\Delta I < 10^{-3}$ bits.  However, HDQ and mSIB achieve  $\Delta I < 10^{-6}$, significantly outperforming the greedy approach of \cite{Tal2011-QuanVardy}.

\section{HDQ Discrete Density Evolution and RCQ Parameter Design}\label{sec: layered}

Discrete density evolution \cite{Chung2000-le} is a technique to  analyze the asymptotic performance of an LDPC ensemble. In this section, we present HDQ discrete density evolution, which is used for designing the quantization thresholds and reconstruction mappings of RCQ decoders and analyzing decoding performance under an RCQ framework. As HDQ discrete density evolution for LDPC decoders with a flooding-schedule has been described thoroughly in our precursor conference paper \cite{Wang2020-RCQ}, this section is focused on HDQ discrete density evolution for LDPC decoders with a  layered schedule. Specifically, this section considers layer-specific msRCQ decoding on QC-LDPC codes.

\subsection{Decoding a Quasi-Cyclic LDPC Code with a Layered Schedule}\label{sec: QC_layer}

QC-LDPC codes are structured LDPC codes characterized by a parity check matrix $H\in \mathbb{F}_2^{(n-k)\times n}$ which consists of square sub-matrices with size $S$, which are either the all-zeros matrix or a cyclic permutation of the identity matrix.   These cyclic permutations  are also called circulants that are represented by $\sigma^i$ to indicate that the rows of the identity matrix are  cyclically shifted  by $i$ positions.  Thus an $M\times U$ \emph{base matrix} $H_\text{p}$ can concisely define a QC-LDPC code, where each element in $H_\text{p}$ is either $\mathbf{0}$ (the all-zeros matrix) or $\sigma^i$ (a circulant). QC-LDPC codes are  perfectly compatible with horizontal layered decoding by partitioning CNs into $M$ layers with each layer containing $S$ consecutive rows. This ensures that each VN connects to at most one CN in each layer.

Denote the $i^{th}$ CN and $j^{th}$ VN by $c_i$ and $v_j$ respectively. Let $u^{(t)}_{c_i\rightarrow v_j}$ be the LLR message from $c_i$ to its neighbor $v_j$ in $t^{th}$ iteration and $l_{v_j}$ be the posterior of $v_j$.  In the $t^{th}$ iteration, a  horizontal-layered \minsum decoder calculates the messages $u^{(t)}_{c_i\rightarrow v_{j'}}$ and updates the posteriors $l_{v_{j'}}$ as follows:
\begin{align}
    {l}_{v_{j'}} &\leftarrow  l_{v_{j'}}- u^{(t-1)}_{c_i\rightarrow v_{j'}}~~\forall j'\in\mathcal{N}(c_i),\label{equ: v-c}
\end{align}
\begin{align}
\begin{split}
        u^{(t)}_{c_i\rightarrow v_{j'}} &=
    \left(\prod_{\tilde{j}\in\mathcal{N}(c_i)/\{{j'}\}}\text{sign}(l_{v_{\tilde{j}}})\right)\\&\times\min_{\tilde{j}\in\mathcal{N}(c_i)/\{{j'}\}}|l_{v_{\tilde{j}}}|,~~\forall j'\in\mathcal{N}(c_i), \label{equ: c-v}
\end{split}
\end{align}
\begin{align}
    l_{v_{j'}} &\leftarrow {l}_{v_{j'}}+u^{(t)}_{c_i \rightarrow v_{j'}} ~~\forall j'\in\mathcal{N}(c_i)\label{equ: posteriot_up}.
\end{align}
$\mathcal{N}(c_i)$ denotes the set of VNs that are neighbors of $c_i$.  For a QC-LDPC code with a long block length, layered decoding is preferable for hardware implementations because  parallel computations of each of \eqref{equ: v-c}, \eqref{equ: c-v}, and \eqref{equ: posteriot_up} exploit the QC-LDPC structure.

\begin{figure}[t] 
    \centering
  \subfloat[ \label{fig: nonlayered_RCQ_str}]{%
      \includegraphics[width=1\linewidth]{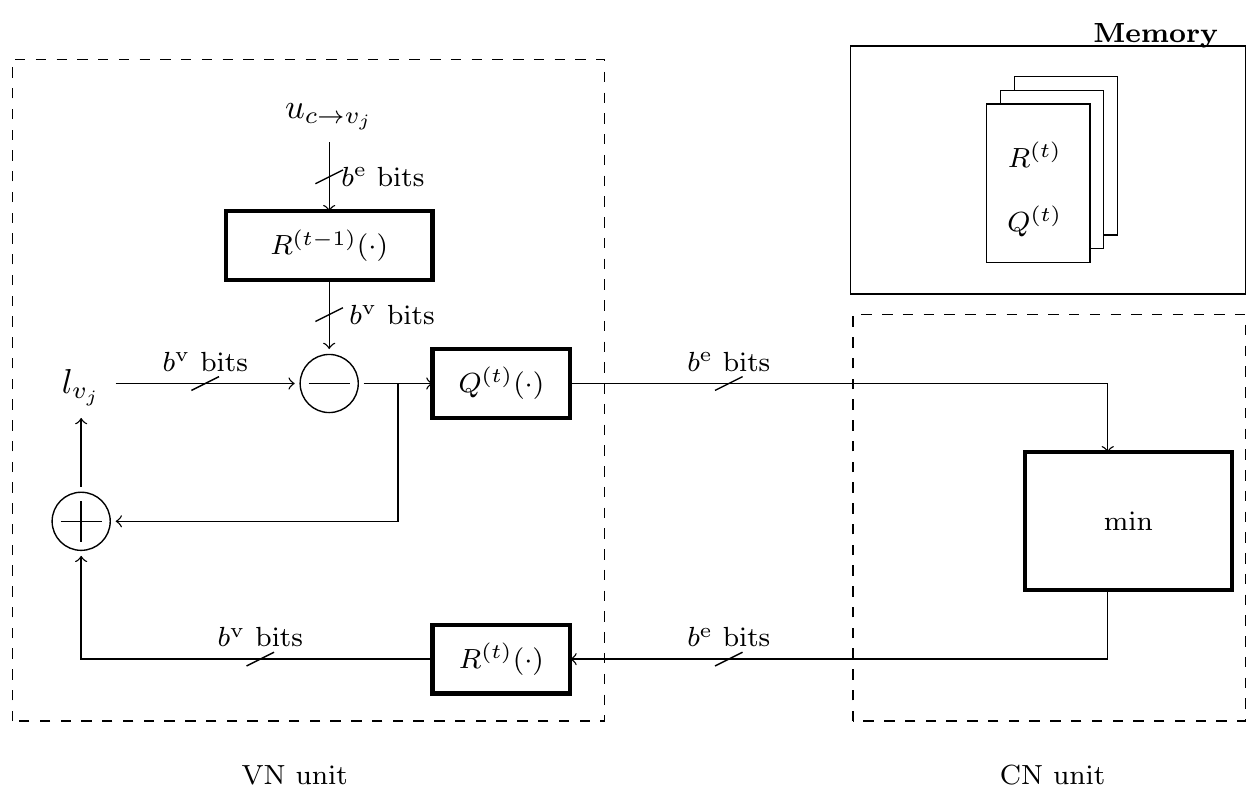}}
\hfill
  \subfloat[\label{fig: layered_RCQ_str}]{%
        \includegraphics[width=1\linewidth]{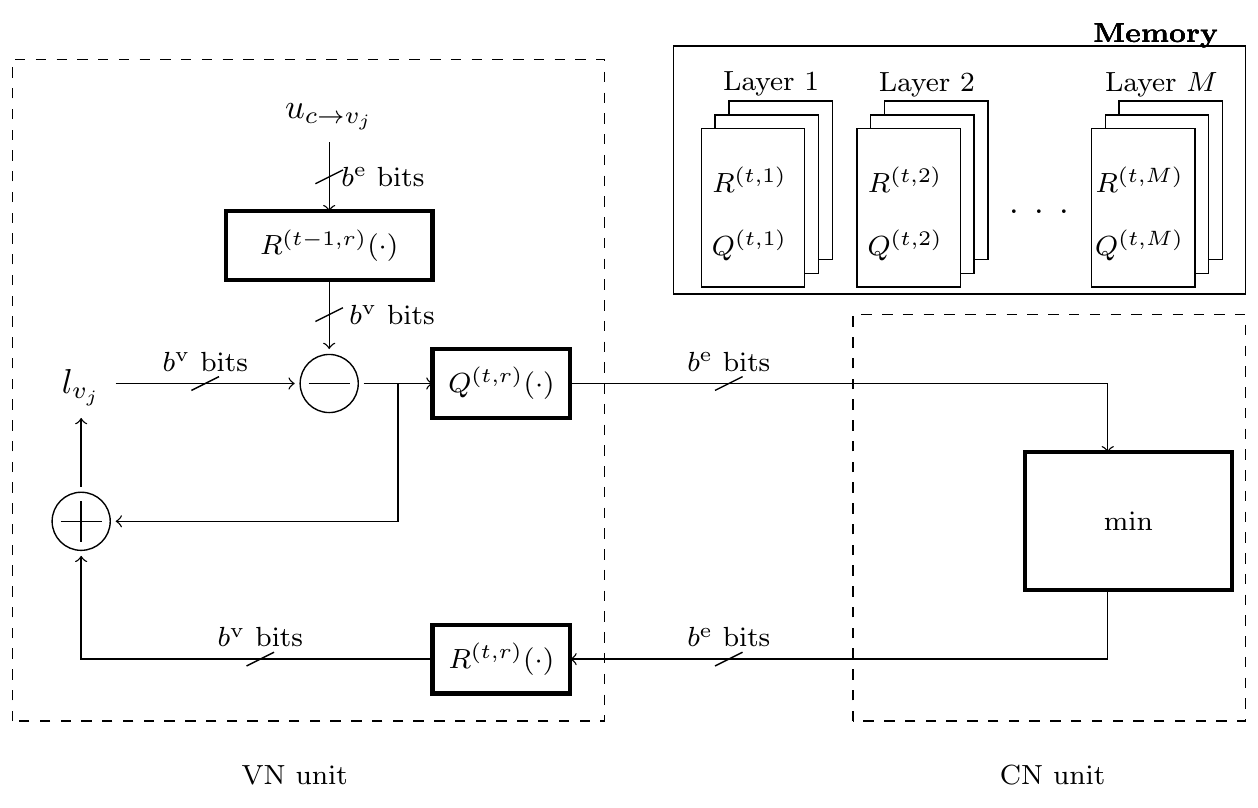}}
  \caption{ Two layered decoders. Fig. (a) uses the same RCQ parameters for each layer as with the  \textit{msRCQ} design for a flooding decoding in \cite{Wang2020-RCQ}. Fig. (b) shows the proposed \textit{layer-specific msRCQ} decoder in \cite{Terrill2021-ec}, which features  separate RCQ parameters for each layer.
  }
  \label{fig: gradient_explosion} 
\end{figure}

\subsection{Representation Mismatch Problem}
The RCQ decoding structure in \cite{Wang2020-RCQ} can be used with a layered schedule as discussed in Sec. \ref{sec: QC_layer}. Fig. \ref{fig: nonlayered_RCQ_str} illustrates the paradigm for an msRCQ decoder with a layered schedule. The $Q_\text{v}^{(t)}$ and $R_\text{v}^{(t)}$ are designed by the HDQ discrete density evolution as in \cite{Wang2020-RCQ}. Even though the msRCQ decoder has better FER performance than the standard \minsum decoder under a flooding schedule\cite{Wang2020-RCQ}, under a layered schedule, msRCQ has worse FER performance than standard \minsum and also requires more iterations.  These performance differences are shown below in Fig.  \ref{fig: 8k_performance} of Sec. \ref{sec: simulation result}. This subsection explains how the performance degradation of the RCQ decoder under the layered schedule is caused by the representation mismatch problem.

Consider a  regular LDPC code defined by a parity check matrix $H$. In iteration $t$, define the PMF between code bit $x$ and external CN messages $u^{(t)}_{c_i \rightarrow v_j}$ as $P^{(t)}_{(c_i,v_j)}(X,D)$, where $X=\{0,1\}$ and $D=\{0,...,2^{b^\mathrm{e}}-1\}$. One underlying assumption of HDQ discrete density evolution is that all CN messages have the same PMF in each iteration, i.e., for any $(c_{i},v_{j})$ and $(c_{i'}, v_{j'})$ that satisfy $H_{i,j}=H_{i',j'}=1$:
\begin{align}
    P^{(t)}_{(c_i,v_j)}(X,D) = P^{(t)}_{(c_{i'},v_{j'})}(X,D).
    \label{equ: Assumption_on_MIM_DDE}
\end{align}
(\ref{equ: Assumption_on_MIM_DDE}) implies that the message indices of different CN have the same LLR representation, i.e.:
\begin{align}
    \log\frac{P^{(t)}_{(c_i,v_j)}(0,d)}{P^{(t)}_{(c_i,v_j)}(1,d)} = \log\frac{P^{(t)}_{(c_{i'},v_{j'})}(0,d)}{P^{(t)}_{(c_{i'},v_{j'})}(1,d)},~   d\in \{0,...,2^{b^\mathrm{e}}-1\}.
    \label{equ: Assumption_on_MIM_DDE2}
\end{align}
 The msRCQ decoder with a flooding schedule obeys (\ref{equ: Assumption_on_MIM_DDE}) and (\ref{equ: Assumption_on_MIM_DDE2})  because the VN messages to calculate different CN messages have the same distribution. Therefore, it is sufficient for a decoder with a flooding schedule  to use the iteration-specific reconstruction function  $R^{(t)}$ for all external CN messages. However, for a decoder with a layered schedule, the VN messages to calculate CN messages from different layers have different distributions. For the decoder with a layered schedule, $l^{(t)}_{v_j\rightarrow c_i}$ is calculated by:
\begin{align}\label{equ: vari_update_2}
\begin{split}
        l^{(t)}_{v_j\rightarrow c_i} = l^{(ch)}_{v_j}&+ \sum_{\{i'|{i'}\in\mathcal{N}(v_j),i'<i\}}u^{(t)}_{c_{i'}\rightarrow v_{j}}\\&+ \sum_{\{i'|{i'}\in\mathcal{N}(v_j),i'>i\}}u^{(t-1)}_{c_{i'}\rightarrow v_{j}},
\end{split}
\end{align}
Unlike a decoder using a flooding schedule, which updates $ l^{(t)}_{v_j\rightarrow c_i}$ only using CN messages in iteration $t-1$, decoders using a layered schedule use messages from both iteration $t-1$ and iteration $t$. The VN messages computed in different layers utilize different proportions of check-to-variable node messages from iterations $t-1$ and $t$.  Since the check-to-variable node messages from different iterations have different reliability distributions, the VN messages from different layers also have different distributions. Therefore \eqref{equ: Assumption_on_MIM_DDE} and \eqref{equ: Assumption_on_MIM_DDE2} no longer hold true, and a single $R^{(t)}(\cdot)$ is insufficient to accurately describe CN messages from different layers. 

In conclusion, the \textit{Representation Mismatch Problem} refers to inappropriately using a single $R^{(t)}$ and single $Q^{(t)}$ for all layers in iteration $t$  of a layered decoding schedule. This issue degrades the decoding performance of layer-scheduled RCQ decoder.  On the other hand, the conventional fixed-point decoders that do not perform coarse non-uniform quantization, such as standard \minsum decoder,  are not affected by the changing the distribution of messages in different layers and hence don't have representation mismatch problem.

\subsection{Layer-Specific RCQ Design}
Based on the analysis in the previous subsection,  $R$ and $Q$ should adapt for the PMF of messages in each layer, in order to solve the representation mismatch problem. This motivates us to propose the layer-specific RCQ decoding structure in this paper, as illustrated in Fig. \ref{fig: layered_RCQ_str}. The key difference between the RCQ decoder and layer-specific RCQ decoder is that layer-specific RCQ designs quantizers and reconstruction mappings for each layer in each iteration.  We use $R^{(t,r)}$ and $Q^{(t,r)}$ to denote the reconstruction mapping and quantizer for decoding iteration $t$ and layer $r$, respectively. 
As illustrated in Fig. \ref{fig: layered_RCQ_str}, layer-specific RCQ specifies $R$ and $Q$ for each layer to handle the issue that messages in different layers have different PMFs. This leads to a significant increase in the required memory because the memory required to store $R^{(t,r)}$ and $Q^{(t,r)}$ is proportional to the product of the number of layers and the number of iterations required for decoding the QC-LDPC code.

Designing $Q^{(t,r)}(\cdot)$ and $R^{(t,r)}(\cdot)$ for layer-specific msRCQ requires the message PMF for each layer in each iteration.  However, HDQ discrete density evolution\cite{Wang2020-RCQ}, which performs density evolution based on ensemble, fails to capture layer-specific information. In this section, we propose a layer-specific HDQ discrete density evolution based on base matrix $H_\text{p}$ of QC-LDPC code. In layer-specific HDQ discrete density evolution, the joint PMF between code bit $X$ and external message $D$ from check/variable nodes are tracked in each layer in each iteration. We use $P^{(t,r)}(X,D^\text{c})$, $X\in\{0,1\}$, $D^\text{c}\in\{0,...,2^{b^\text{e}}-1\}$ to represent the joint PMF between code bit and CN message in layer $m$ and iteration $t$. Similarly, VN messages are denoted by $P^{(t,r)}(X,D^\text{v})$.
\subsubsection{Initialization}
For an AWGN channel with noise variance $\sigma^2$, the LLR of channel observation $y$ is $l=\frac{2}{\sigma^2}y$.
For the msRCQ decoder with bit width $(b^\text{e},b^\text{v})$, the continuous channel LLR input is uniformly quantized into $2^{b^\text{v}}$ regions.  Each quantization region has a true log likelihood ratio, which we refer to as $l_d$, so that we have an alphabet of $b^\text{v}$ real-valued log likelihood ratios ${\cal D}^{\text{ch}}=\{l_0,...,l_{2^{b^{\text{v}}}-1}\}$.  Using these values, the joint PMF between the code bit $X$ and channel LLR message $D^{\text{ch}} \in \{0,...,{2^{b^{\text{v}}}-1}\}$ is:
\begin{align}\label{equ: chan_prob}
    P_{XD^{\text{ch}}}(x,d) &= P_D(d) \frac{e^{(1-x)l_d}}{e^{l_d}+1}, ~X\in \{0,1\},\  l_d\in {\cal D}^{\text{ch}}\,.
\end{align}
The distribution $P_{XD^{\text{ch}}}(x,d)$ is used for the HDQ discrete density evolution design.  The actual decoder does not use the real-valued likelihoods $l_d$ but rather uses $b^{\text{v}}$-bit channel LLRs obtained by uniformly quantizing continuous channel LLR values.
\subsubsection{Variable Nodes PMF Calculation}
Given a base matrix $H_\text{p}$, with entry $H_\text{p}(r,c)$ at row $r$ and column $c$, define the sets of active rows $\mathcal{R}(c)$ for a specified column $c$ and active columns $ \mathcal{C}(r)$ for a specified row $r$ as follows:
\begin{align}
\mathcal{R}(c) = \{r| H_\text{p}(r,c)\neq0\}
,\quad 
\mathcal{C}(r)= \{c| H_\text{p}(r,c)\neq0\}.
\end{align}
In iteration $t$ and layer $r$, consider the joint PMF between a code bit $X$ corresponding to a VN in the circulant $H_\text{p}(r,c)$ and the vector $\mathbf{D}$, which includes the channel message $D^{\text{ch}}$ for $X$ and the check node messages $D^{\text{c}}$ incident to that VN.  This PMF is calculated by:
\begin{align}
\begin{split}
        P_\text{v}^{(t,r,c)}(X,\mathbf{D})&=P(X,D^{\text{ch}})\boxdot\left(\boxdot_{\substack{k\in\mathcal{R}(c) \\ k<r}}P^{(t,k)}(X,D^{\text{c}})\right) \\&\boxdot\left(\boxdot_{\substack{k\in\mathcal{R}(c) \\ k>r}}P^{(t-1,k)}(X,D^{\text{c}})\right), 
\end{split}
    \label{eq: vari_update}
\end{align}
$\boxdot$ is defined as follows: 
\begin{align}
    P(x,[d_1,d_2])&=P({X_1,D_1})\boxdot P({X_2,D_2})\\&\triangleq\frac{1}{P_X(x)}P_{X_1D_1}(x,d_1)P_{X_2D_2}(x,d_2),
\end{align}
$ x\in\{0,1\},~d_1,d_2\in\{0,...,2^{b^\mathrm{e}}-1\}$. When $|\mathcal{R}(c)|$ is large, the alphabet ${\mathcal D}$ of possible input message vectors $\mathbf{D}$ is large with $|{\mathcal D}|=2^{b^\mathrm{v}+(|\mathcal{R}(c)|-1)b^\mathrm{e}}$.  To manage the complexity of HDQ discrete density evolution, message vectors $\mathbf{D}$ with similar log likelihoods are clustered via  one-step-annealing as in \cite{Wang2020-RCQ} for (\Ref{eq: vari_update}).

The layer-specific msRCQ decoder uses layer-specific parameters, and for each layer the marginal distribution on the computed variable node messages will be distinct. The marginal distribution used by HDQ at layer $r$ is computed as follows:
\begin{align}\label{eq: vari_total}
    \Tilde{P}_\text{v}^{(t,r)} = \left\{\frac{1}{|\mathcal{C}(r)|}P_\text{v}^{(t,r,c)}(X,\mathbf{D})\mid c\in\mathcal{C}(r)\right\}\,
\end{align}
 where ${P}^{(t,r)}(X,D^v)$ and $Q^{(t,r)}(\cdot)$ can be obtained by quantizing  $\Tilde{P}_\text{v}^{(t,r)}$ using HDQ:
\begin{align}\label{eq: vari_hdq}
    \left[P^{(t,r)}(X,D^{\text{v}}),Q^{(t,r)}(\cdot) \right] = \texttt{HDQ}\left(\Tilde{P}_\text{v}^{(t,r)}, 2^{b^\mathrm{e}}\right),
\end{align}
where \texttt{HDQ} is defined as a function that realizes $b^e$-bit HDQ on  $\Tilde{P}_\text{v}^{(t,r)}$ and generates $P^{(t,r)}(X,D^{\text{v}})$ and $Q^{(t,r)}$ as outputs. Note that (33) and (34) realize implicit message alignment in \cite{Stark2021-ai} such that the internal messages from  any  $c\in\mathcal{C}(r)$  use same set of thresholds for quantization and the same external messages from any $c\in\mathcal{C}(r)$ have same LLR interpretations, regardless of node degree.

\subsubsection{Check Nodes PMF Calculation}
Let $l^{(t,r)}_\text{v}{(d)}$ be the LLR of external VN message $d$ in layer $r$ and iteration $t$.  As an LLR, this CN input  $l^{(t,r)}_\text{v}{(d)}$ has the following meaning: 
\begin{align}
    l_\text{v}^{(t,r)}(d) = \log \frac{P_{XD^\text{v}}^{(t,r)}(0,d)}{P_{XD^\text{v}}^{(t,r)}(1,d)},\quad d=0,...,2^{b^{e}}-1.
\end{align}
Given input messages $d_1,d_2\in {\mathcal D}^\text{v}$, the CN min operation produces the following output:
\begin{align}\label{equ: label_min}
\begin{split}
     l_{\text{MS}}^{\text{out}} =&\min\left(|l_\text{v}^{(t,r)}(d_1)|,|l_\text{v}^{(t,r)}(d_2))|\right)\\&\times \text{sgn}(l_\text{v}^{(t,r)}(d_1))\times\text{sgn}(l_\text{v}^{(t,r)}(d_2)). 
\end{split}
\end{align}
Under the symmetry assumption, there is a $d^{\text{out}}\in {\mathcal D}^\text{v}$ that has the LLR computed as $l_{\text{MS}}^{\text{out}}$:
\begin{align}\label{equ: label_min_2}
    l_{\text{MS}}^{\text{out}}=\log \frac{P^{(t,r)}_{XD^\text{v}}(0,d^{\text{out}})}{P^{(t,r)}_{XD^\text{v}}(1,d^{\text{out}})}.
\end{align}

Define the follow function: 
\begin{align}\label{equ: label-min}
    d^\text{out} = \texttt{MS}(d_1,d_2),
\end{align}
 where $d^{\text{out}},d_1,d_2\in \mathcal{D}^\mathrm{v} $. \eqref{equ: label-min} holds if and only if \eqref{equ: label_min} and \eqref{equ: label_min_2} and are both satisfied.

Define the binary operation $\circledast$ by:
\begin{align}
          \Tilde{P}_{XD}(x,d) &= P(X_1,D_1)\circledast P(X_2,D_2)\\&\triangleq\sum_{\substack{d_1,d_2:\texttt{MS}(d_1,d_2)=d \\ x_1,x_2: x_1\bigoplus x_2 = x}}P_{X_1D_1}(x_1,d_1)P_{X_2D_2}(x_2,d_2)\label{equ: msrcq-c}.
\end{align}

The joint PMF between code bit and external CN message in layer $r$ and iteration $t$ can be updated by:
\begin{align}
 P^{(t,r)}(X,D^\text{c}) &=P^{(t,r)}(X,D^\text{v})\circledast ...\circledast P^{(t,r)}(X,D^\text{v})\\&\triangleq P^{(t,r)}(X,D^\text{v})^{\circledast(|\mathcal{C}(r)|-1)}. \label{equ: check_opt_prob}
\end{align}
$R^{(t,r)}(\cdot)$ can be directly computed using $P^{(t,r)}(X,D^\text{c})$:
\begin{align}
        R^{(t,r)}(d) = \log \frac{P_{XD^\text{c}}^{(t,r)}(0,d)}{p_{XD^\text{c}}^{(t,r)}(1,d)},~d\in\{0,...,2^{b^\text{e}}-1\}.
        \label{eq: r_v}
\end{align}


\subsection{Threshold}
At any specified $\frac{E_b}{N_o}$, layer-specific HDQ discrete density evolution constructs the $R^{(t,r)}(\cdot)$ and $Q^{(t,r)}(\cdot)$ functions for each layer $r$ at each iteration $t$ and also computes the mutual information $I^{(t,r)}\left(\frac{E_b}{N_o}\right)$ between a code bit and its corresponding variable node message in each layer $r$ at each iteration $t$. An important design question is which value of $\frac{E_b}{N_o}$ to use to construct the $R^{(t,r)}(\cdot)$ and $Q^{(t,r)}(\cdot)$ functions implemented at the decoder, which necessarily will work over a range of $\frac{E_b}{N_o}$ values in practice.  Define the threshold of a layer-specific RCQ decoder given a base matrix with $M$ layers and  maximum number of decoding iterations $I_T$ as: 
\begin{align}
    \frac{E_b}{N_o}^*=\inf \left\{\frac{E_b}{N_o}:  I^{(I_T,r)}\left(\frac{E_b}{N_o}\right)> 1-\epsilon,\forall  r\in[1,M] \right\},
\end{align}
i.e., $\frac{E_b}{N_o}^*$ is the smallest $\frac{E_b}{N_o}$ that achieves a mutual information between the  code bit and the external message that is greater that $1-\epsilon$ for each layer. Our simulation results show that  $\frac{E_b}{N_o}^*$ for $\epsilon = 10^{-4}$ produced $R^{(t,r)}(\cdot)$ and $Q^{(t,r)}(\cdot)$ functions that deliver excellent FER performance across a wide $\frac{E_b}{N_o}$ range. 
\begin{figure}[t] 
    \centering
  \subfloat[Decoders with floating point messages\label{fig: 80211_1}]{%
       \includegraphics[width=1\linewidth]{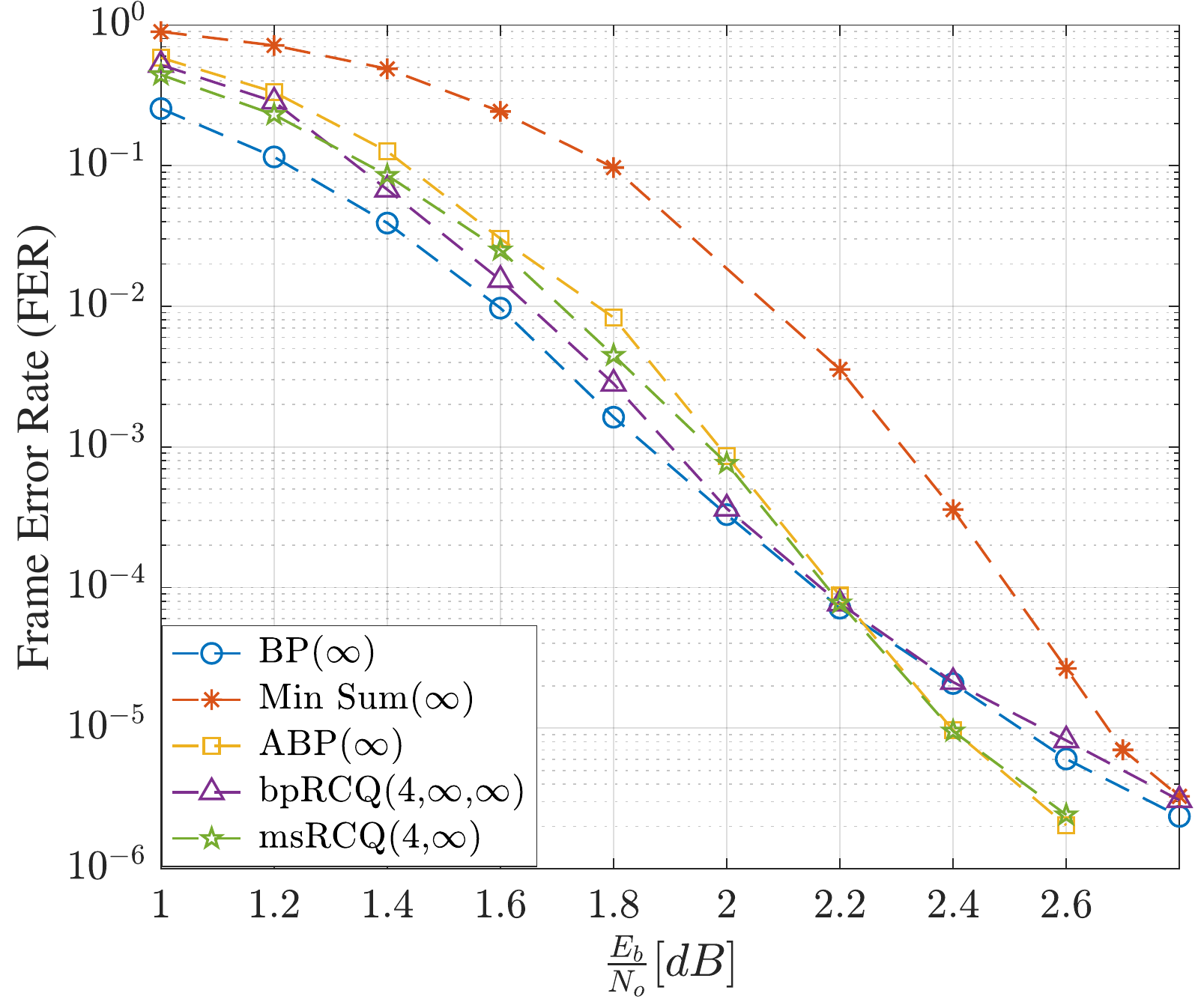}}
    \hfill
  \subfloat[Decoders with fixed point messages\label{fig: 80211_3}]{%
        \includegraphics[width=1\linewidth]{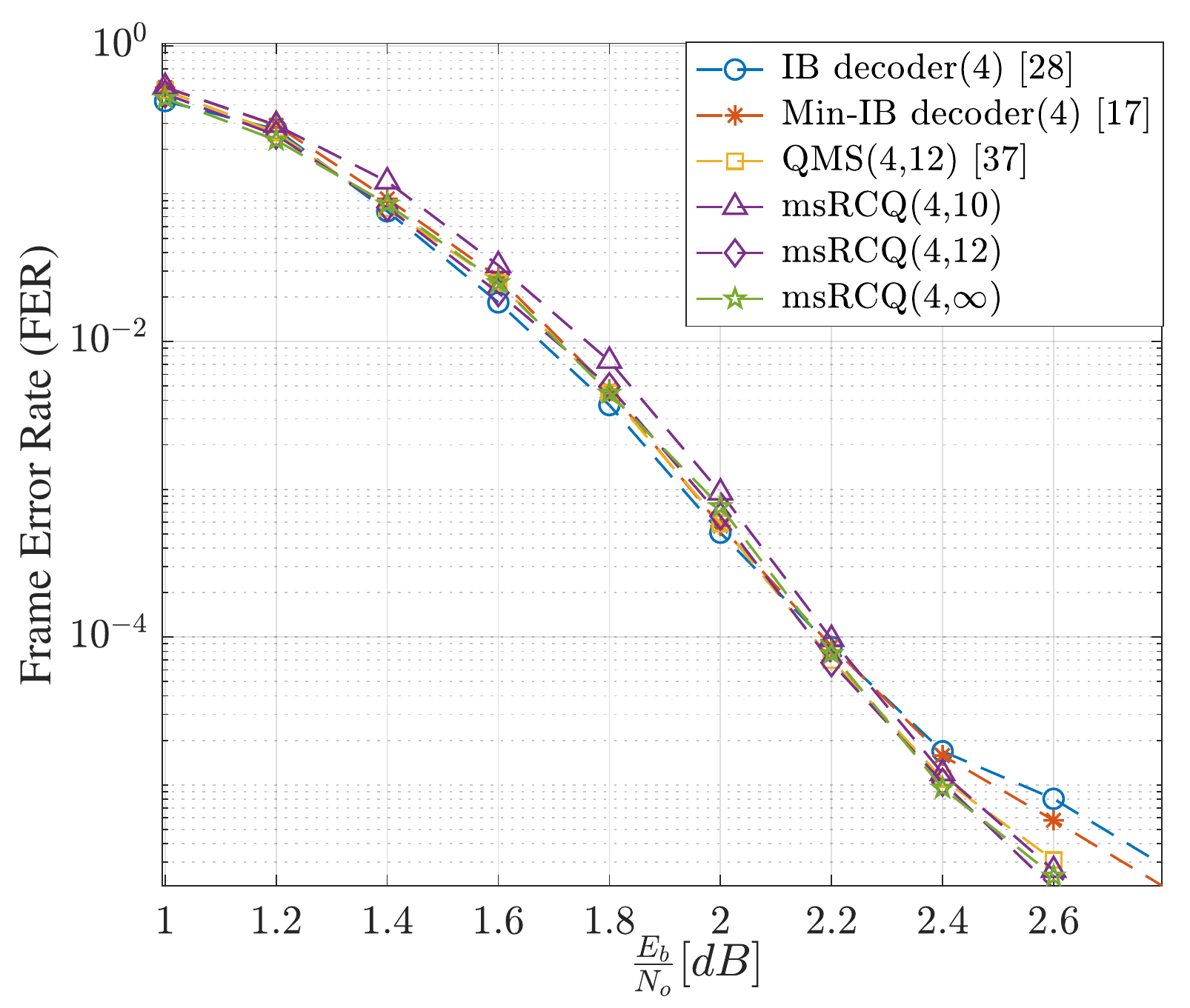}}
  \caption{Fig. (a): FER performance of 4-bit msRCQ and bpRCQ decoders with floating point message representations use at the VNs.  Fig. (b):FER performance of fixed point 4-bit msRCQ decoders, compared with other non-uniform quantization decoders. }
\end{figure}

\section{Simulation Result and Discussion}\label{sec: simulation result}
This section presents RCQ and layer-specific RCQ decoder designs for two example LDPC codes and compares their FER performance  with existing conventional decoders such as BP, \minsum, and state-of-the-art non-uniform decoders, such as an IB decoder. All decoders are simulated using the AWGN channel, and at least 100 frame errors are collected for each point.  We also compare hardware requirements for an example LDPC code.
\subsection{IEEE 802.11 Standard LDPC Code}
We first investigate the FER performance of RCQ decoders with a flooding schedule using an IEEE 802.11n standard LDPC code taken from \cite{80211spec}. This code has $n=1296$, $k=648$, 
and the maximum number of decoding iterations was set to 50.

Fig. \ref{fig: 80211_1} shows the FER curves of  4-bit bpRCQ and msRCQ decoder with floating-point internal messages, i.e., {bpRCQ}(4,$\infty$,$\infty$) and {msRCQ}(4,$\infty$), respectively . The notation of $\infty$ represents floating-point message representation. Denote floating point BP nad Min Sum by BP($\infty$) and Min Sum($
\infty$), respectively.  The 4-bit bpRCQ decoder has at most 0.1 $dB$ degradation compared with the floating-point BP decoder, and outperforms floating-point BP at high $\frac{E_b}{N_o}$.  The 4-bit msRCQ performs better than conventional \minsum and even surpasses BP at high $\frac{E_b}{N_o}$. The lower error floor of msRCQ decoder as compared to standard BP follows from the slower message magnitude convergence rate as compared to standard BP. This is similar to improved error floors achieved by the averaged BP (ABP) [35], which  decreases the rate of increase of message magnitudes by averaging the posteriors  $l_v^{(t)}$ in consecutive iterations. As shown in Fig. \ref{fig: 80211_1}, ABP also delivers a lower error floor than standard BP.

 The slow magnitude convergence rate of msRCQ decoder can be explained as follows.  For conventional \minsum decoder, the magnitude of each check node message is always equal to the magnitude of an input variable node message for that CN.  This is not true for the msRCQ decoder. msRCQ compares the relative LLR meanings of input messages and returns an external message by implementing the min operation. However, the external message is then reconstructed at the VN to an internal message magnitude that is in general different from the message magnitudes that were received by the neighboring CN.

For the example of a degree-3 CN, \eqref{equ: ave_msrcq} computes the likelihood associated with a message $l_t$ that is outputted from the min operation applied to the other two input messages indexed by $i$ and $j$:
\begin{align}\label{equ: ave_msrcq}
    l_{t}&=\log\frac{\sum_{\{(i,j)|t=MS(i,j)\}}P(0,i)P(0,j)+P(1,i)P(1,j)}{\sum_{\{(i,j)|t=MS(i,j)\}}P(1,i)P(0,j)+P(0,i)P(1,j)}. 
\end{align}
Note that the boxplus operation is computed as follows :
\begin{align}
   l_i\boxplus l_j &=\log \frac{P(0,i)P(0,j)+P(1,i)P(1,j)}{P(0,i)P(1,j)+P(1,i)P(0,j)}\label{equ: box_plus_p}.
\end{align}
Comparing with \eqref{equ: box_plus_p}, it can be seen that \eqref{equ: ave_msrcq} applies the boxplus operation to the probability of the group of messages that share same value for $MS(i,j)$. Applying the boxplus operation to the {\em group} of messages produces a value that lies between the extremes of the messages produced by individual boxplus operations.  This grouping process lowers the maximum output magnitude and therefore decreases the message magnitude growth rate in an iterative decoding process. 
\begin{figure}
	\centering
	\includegraphics[width=18pc]{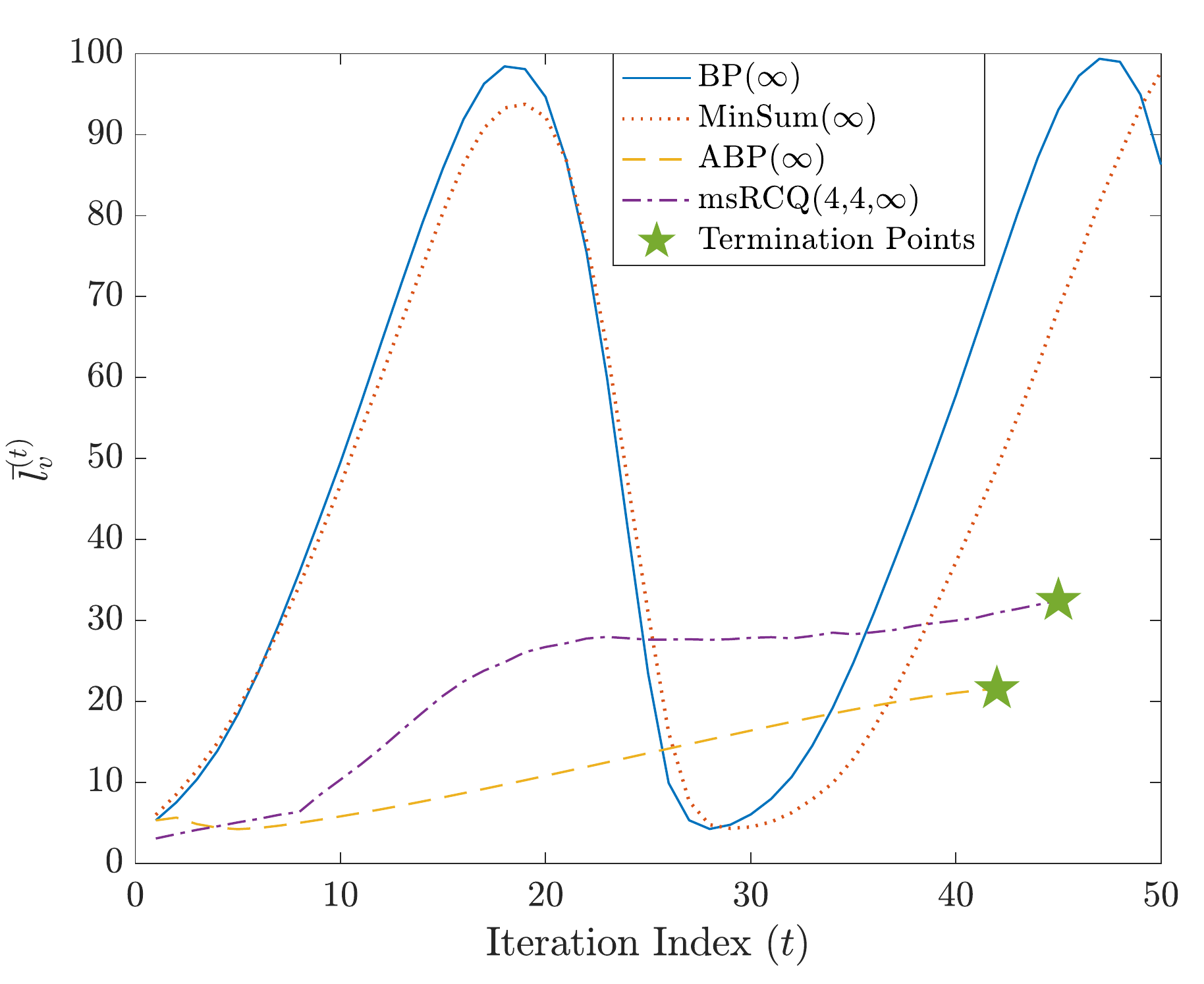}
	\caption{Average magnitudes of $l^{(t)}_v$ vs. iteration for BP, ABP, \minsum and msRCQ for Fig.  6a simulation at  $\frac{E_b}{N_o}=2.6$ dB.
	}
	\label{fig: 80211_2}
\end{figure}
As noted in \cite{ABP}, a possible indicator of the emergence of error trapping sets may be a sudden magnitude change in the values of certain variable node messages, or fast convergence to an unreliable estimate. Therefore, slowing down the convergence rate of VN messages can decrease the frequency of trapping set events. Both msRCQ decoder and A-BP in [35] reduce the the convergence rate of VN messages and hence deliver a lower error floor. 

The effect of averaging can be seen in Fig. \ref{fig: 80211_2}, which gives the average magnitude of $l^{(t)}_v$ for four decoders with a noise-corrupted all-zero codeword at $\frac{E_b}{N_o}=2.6$ dB as the input. The oscillation pattern of the BP decoder has been reported and discussed in \cite{ABP}.  As shown in Fig. \ref{fig: 80211_1}, ABP also outperforms belief propagation when $\frac{E_b}{N_o}$ is high. 

Fig. \ref{fig: 80211_3} compares msRCQ(4,10) with other non-uniform quantization LDPC decoders. Simulation results show that both IB\cite{Lewandowsky2018-IBRegular} and Min-IB\cite{Meidlinger2017-MINIBIRR} decoders exhibit an error floor after $2.40dB$. The MIM-QMS\cite{kang2020generalized} decoder has a similar decoding structure to msRCQ. Note that MIM-QMS requires the determination of the internal bit width used by the VNs before designing quantization and reconstruction parameters, so reducing the bit width of VNs requires another design cycle. 
In contrast, for the purposes of HDQ discrete density evolution design process, msRCQ assumes that the internal VN  messages are real-valued. This assumption is an approximation since the internal VN messages will have finite precision in practical implementations. During actual decoding, the reconstruction operation $R(\cdot)$ produces a high-precision representation for use in computations at the VN. We found that assuming real-valued internal messages in the design process introduces negligible loss for practical internal message sizes while greatly simplifying the design. Our simulation results in \ref{fig: 80211_3} confirm that high precision internal messages have FER performance that is very close to real-valued internal messages.
The RCQ decoder has more efficient memory usage than LUT-based decoders. For the investigated non-uniform LDPC code, 4-bit IB and 4-bit Min-IB require 14.43k and 10.24k bits, respectively, for storing LUTs per iteration, whereas msRCQ(4,12) and msRCQ(4,10) require 165 bits and 135 bits only.

\subsection{(9472, 8192) QC-LDPC code}
In this section we consider a rate-0.8649 quasi-regular LDPC code, with all VNs having degree 4 and CNs having degree 29 and 30, as might be used in a flash memory controller. We study this (9472, 8192) QC-LDPC code using various decoders with a \textit{layered schedule}. The layer number of the investigated LDPC code is 10.  
\begin{figure}[t] 
    \centering
  \subfloat[\label{fig: 8k_FER}]{%
       \includegraphics[width=1\linewidth]{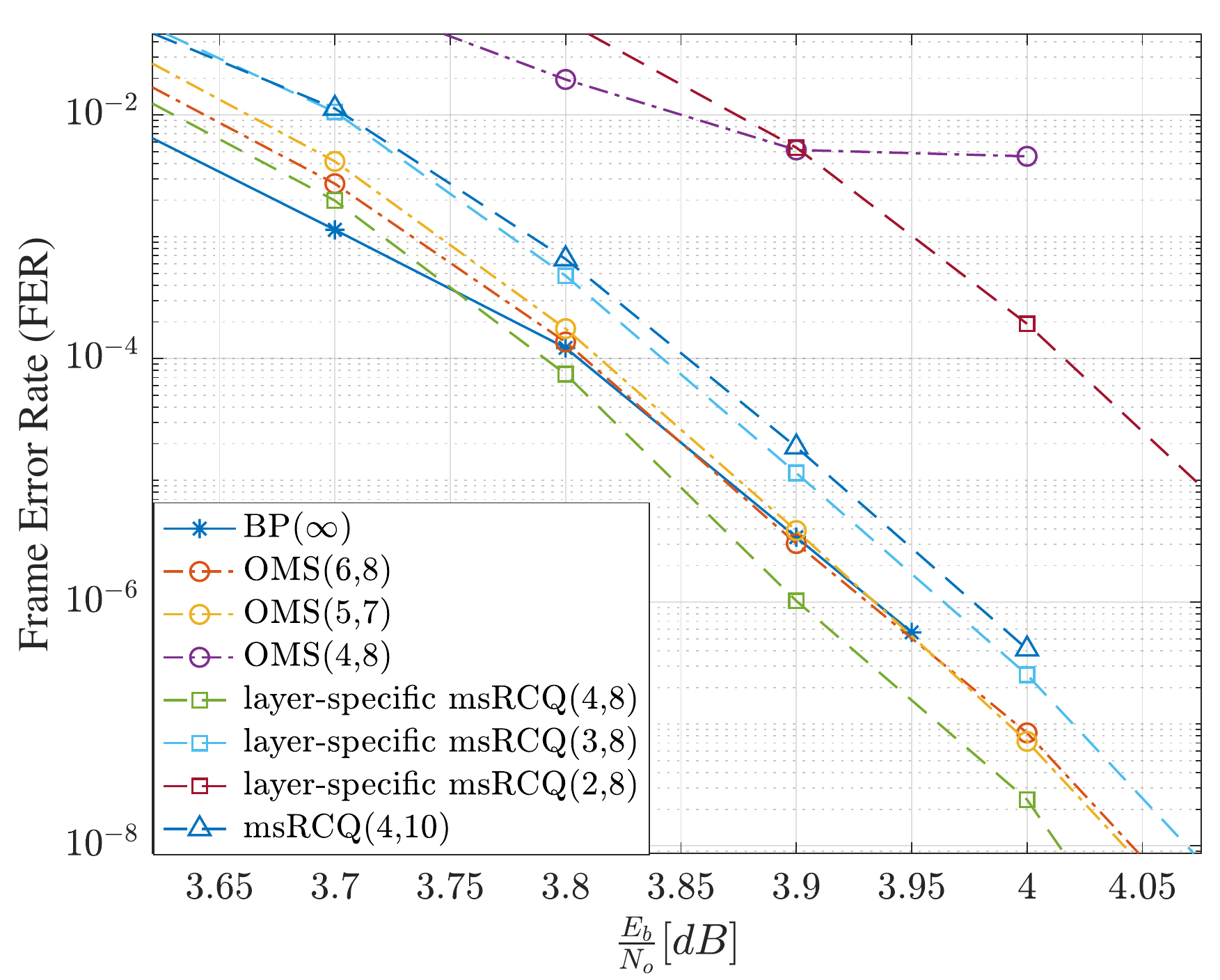}}
    \hfill
  \subfloat[\label{fig: 8k_ADIT}]{%
        \includegraphics[width=1\linewidth]{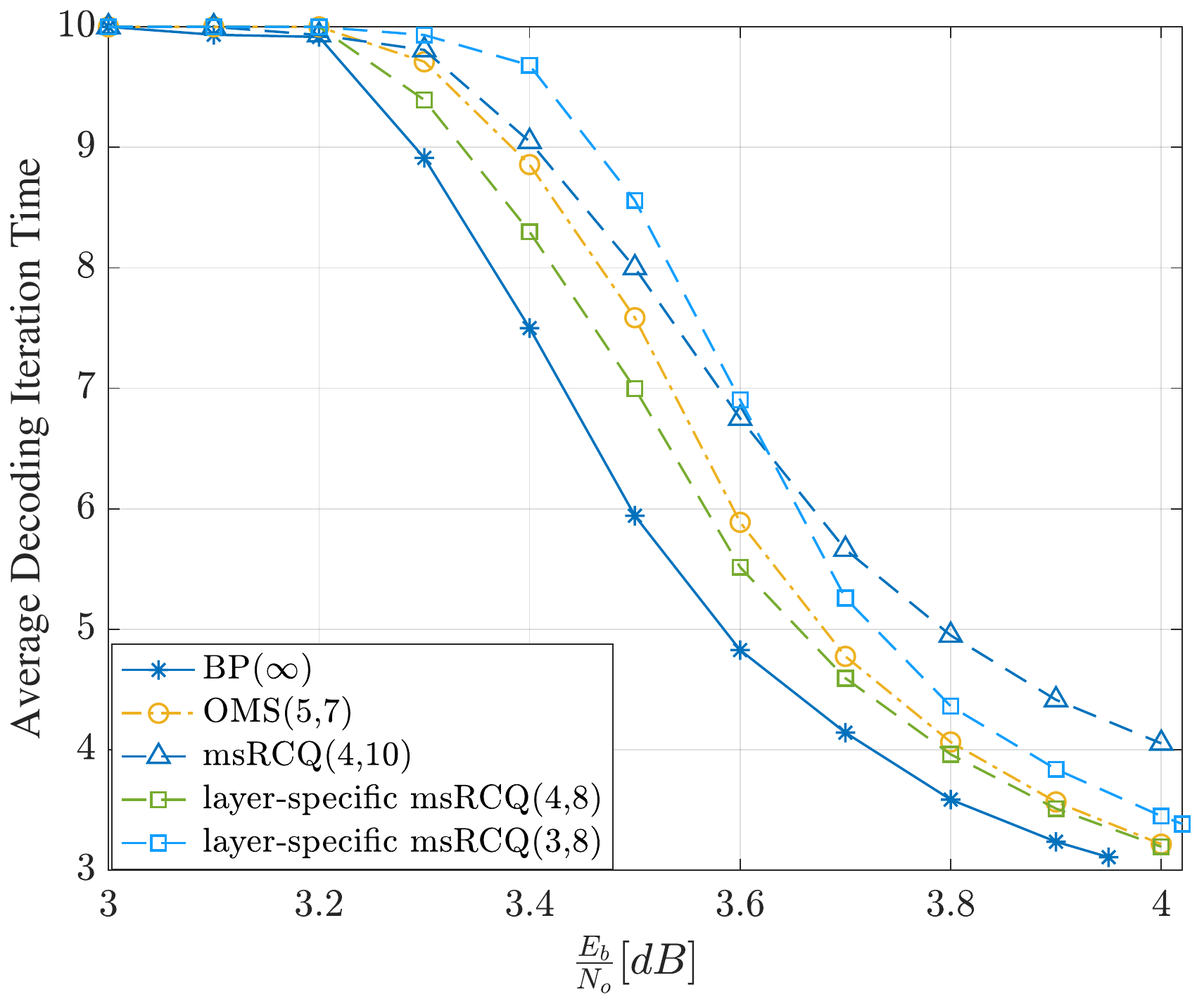}}
  \caption{Fig. (a): FER performance of fixed point L-msRCQ decoders for (9472, 8192) LDPC code.  Fig.  (b): FER performance of fixed point L-msRCQ decoders for (9472, 8192) LDPC code. } \label{fig: 8k_performance}
\end{figure}

Fig. \ref{fig: 8k_FER} shows the FER curves of various decoders. The maximum number of decoding iterations of all studied decoders is 10. The layer-specific msRCQ(4,8) outperforms msRCQ(4,10) by 0.04 dB, which shows the benefit of optimizing layer and iteration specific RCQ parameters.  
The layer-specific msRCQ(3,8) delivers similar decoding performance to msRCQ(4,10). The decoding performance of 2-bit layer-specific msRCQ has a 0.2 dB degradation compared with the 4-bit  layer-specific msRCQ decoder. 
Fig. \ref{fig: 8k_FER} also shows a fixed point offset \minsum (OMS) decoder with offset factor 0.5. At a FER of $10^{-8}$, OMS(6,8) and OMS(5,7) outperform layer-specific msRCQ(3,8) by 0.02 dB, yet are inferior to layer-specific msRCQ(4,8) by 0.02 dB.  
\begin{table*}[t]
\centering
\caption{\label{tab: hardware_usage} Hardware Usage of Various Decoding Structure for (9472,8192) QC-LDPC Code }
\begin{tabular}{|c|c|c|c|c|}
\hline
Decoding Structure & LUTs                       & Registers                  & BRAMS                        & Routed Nets                \\ \hline
OMS(5,7) (baseline)          & 21127                      & 12966                     & 17                    & 29202                      \\ \hline
layer-specific RCQ(4,8)           & 20355(\textcolor{mygreen}{$\downarrow 3.6\%$} ) & 13967(\textcolor{red}{$\uparrow 7.0\%$}) & 17.5(\textcolor{red}{$\uparrow .03\%$}) & 28916(\textcolor{mygreen}{$\downarrow 1\%$}) \\ \hline
layer-specific RCQ(3,8)           & 17865(\textcolor{mygreen}{$\downarrow 15.4\%$}) & 12098(\textcolor{mygreen}{$\downarrow 6.7\%$}) & 17(\textcolor{blue}{$-$})                &  25332\textcolor{mygreen}{($\downarrow 13.3\%$}) \\ \hline
\end{tabular}
\end{table*}
Fig. \ref{fig: 8k_ADIT} shows the average decoding iteration times for some of the decoders studied in Fig. \ref{fig: 8k_FER}. At high $\frac{E_b}{N_o}$, the msRCQ(4,10) decoder requires the largest average number of iterations to complete decoding. On the other hand, layer-specific msRCQ(4,8) has a similar decoding iteration time to OMS(5,7) and BP($\infty$) in this region. Layer-specific msRCQ(3,8) requires a slightly higher average number of iterations than layer-specific msRCQ(4,8) and OMS(5,7). 

We implemented OMS and layer-specific msRCQ decoders with different bit widths on the programmable logic of a Xilinx Zynq UltraScale+ MPSoC device for comparison. Each design meets timing with a 500 MHz clock. The broadcast method described in \cite{Terrill2021-ec} is used for RCQ design. 
Table \ref{tab: hardware_usage} summarizes the hardware usage of each decoder. 
Simulation result shows that layer-specific msRCQ(4,8) has a similar hardware usage with OMS(5,7), and layer-specific msRCQ(3,8) has more than a $10\%$ reduction in LUTs and routed nets and more than a $6\%$ reduction in registers, compared with OMS(5,7).

\section{Conclusion}\label{sec: conclusion}
This paper investigates the decoding performance and resource usage of RCQ decoders. For decoders using the flooding schedule, simulation results on an IEEE 802.11 LDPC code show that a 4-bit msRCQ decoder has a better decoding performance than LUT based decoders, such as IB decoders or Min-IB decoders, with significantly fewer parameters to be stored. It also surpasses belief propagation in the high $\frac{E_b}{N_o}$ region because a slower message convergence rate avoids trapping sets. For decoders using the layered schedule, conventional RCQ design leads to a degradation of FER performance and higher average decoding iteration time. Designing a layer-specific RCQ decoder, which updates parameters in each layer and iteration, improves the performance of a conventional RCQ decoder under a layered schedule.
Layer-specific HDQ discrete density evolution is proposed to design parameters for RCQ decoders with a layered schedule. FPGA implementations of RCQ decoders are used to compare the resource requirements of the decoders studied in this paper. 
Simulation results for a (9472, 8192) QC LDPC code show that a layer-specific \minsum RCQ decoder with 3-bit messages achieves a more than $10\%$ reduction in LUTs and routed nets and a more than $6\%$ register reduction while maintaining comparable decoding performance, compared to a 5-bit offset \minsum decoder.

\bibliographystyle{IEEEtran}
\bibliography{IEEEabrv,GallagerThesis,RCQ_TCCOM}

\end{document}